\documentclass{aa}  

\usepackage{graphicx}
\usepackage{txfonts}
\usepackage[colorlinks=true,citecolor=blue,linkcolor=blue]{hyperref}
\begin{document}

   \title{Self-generated ultraviolet radiation in molecular shock waves}
   \titlerunning{Self-generated UV in molecular shocks}
   \subtitle{I. Effects of Lyman~$\alpha$, Lyman~$\beta$, and two-photon continuum}
   \author{A. Lehmann\inst{1}, B. Godard\inst{1,2}, G. Pineau des For\^{e}ts\inst{2,3}, \and E. Falgarone\inst{1}}
   \authorrunning{Lehmann, Godard, Pineau des For\^{e}ts \& Falgarone}

   \institute{Laboratoire de Physique de l'ENS, ENS, Universit\'{e} PSL, CNRS, Sorbonne Universit\'{e}, Universit\'{e} Paris-Diderot, Paris, France
	\email{andrew.lehmann@ens.fr}
   		\and
   LERMA, Observatoire de Paris, PSL Research Univ., CNRS, Sorbonne Univ., 75014 Paris, France
   		\and
   Institut d’Astrophysique Spatiale, CNRS, Université Paris-Saclay, Bât. 121, 91405 Orsay Cedex,
France 
   }
   \date{Submitted June 12, 2020}

 
  \abstract
  
\abstract{Shocks are ubiquitous in the interstellar and intergalactic media, where their chemical and radiative signatures reveal the physical conditions in which they arise. Detailed astrochemical models of shocks at all velocities are necessary to understand the physics of many environments including protostellar outflows, supernova remnants, and galactic outflows.}
{We present an accurate treatment of the self-generated ultraviolet (UV) radiation in models of intermediate velocity ($V_S = 25 - 60$ km/s), stationary, weakly magnetised, J-type, molecular shocks. We show how these UV photons modify the structure and chemical properties of shocks and quantify how the initial mechanical energy is reprocessed into line emission.}
{We develop an iterative scheme to calculate the self-consistent UV radiation field produced by molecular shocks. The shock solutions computed with the Paris-Durham shock code are post-processed using a multi-level accelerated $\Lambda$-iteration radiative transfer algorithm to compute Lyman $\alpha$, Lyman $\beta$, and two-photon continuum emission. The subsequent impacts of these photons on the ionisation and dissociation of key atomic and molecular species as well as on the heating by the photoelectric effect are calculated by taking the wavelength dependent interaction cross-sections and the fluid velocity profile into account. This leads to an accurate description of the propagation of photons and the thermochemical properties of the gas in both the postshock region and in the material ahead of the shock called the radiative precursor. With this new treatment, we analyse a grid of shock models with velocities in the range $V_S=25-60$ km/s, propagating in dense ($n_{\rm H} \geqslant 10^4$ cm$^{-3}$) and shielded gas.}
{Self-absorption traps Ly$\alpha$ photons in a small region in the shock, though a large fraction of this emission escapes by scattering into the line wings. We find a critical velocity $V_S\sim 30$ km/s above which shocks generate Ly$\alpha$ emission with a photon flux exceeding the flux of the standard interstellar radiation field. The escaping photons generate a warm slab of gas ($T\sim 100$ K) ahead of the shock front as well as pre-ionising C and S. Intermediate velocity molecular shocks are traced by bright emission of many atomic fine structure (e.g. O and S) and metastable (e.g. O and C) lines, substantive molecular emission (e.g. H$_2$, OH, and CO), enhanced column densities of several species including CH$^+$ and HCO$^+$, as well as a severe destruction of H$_2$O.As much as 13-21\% of the initial kinetic energy of the shock escapes in Ly$\alpha$ and Ly$\beta$ photons if the dust opacity in the radiative precursor allows it.}
{A rich molecular emission is produced by interstellar shocks regardless of the input mechanical energy. Atomic and molecular lines reprocess the quasi totality of the kinetic energy, allowing for the connection of observable emission to the driving source for that emission.}

   \keywords{shock waves --
             radiative transfer --
             ISM: kinematics and dynamics --
             ISM: molecules --
             ISM: atoms --
             methods: numerical
             }

   \maketitle
%
\section{Introduction}

Shocks are the fingerprints of the dynamical state and evolution of the interstellar medium. From protostellar outflows and supernova explosions to galactic outflows, massive amounts of mechanical energy are injected at supersonic velocities into the entire range of galactic scales and interstellar phases \citep[e.g.][]{Raymond2020}. Such flows inevitably form shocks which immediately alter the state of the gas via compression and viscous heating \citep{draine_1993}, but they also give rise to a turbulent cascade and the generation of lower velocity shocks \citep[e.g.][]{Elmegreen2004}. By reprocessing the initial kinetic energy into atomic and molecular lines, shocks produce invaluable tracers which carry information on the physical conditions of the shocked gas and on the source itself, including its energy budget, lifetime, and mass ejection rate (e.g. \citealt{Reach2005,McDonald2012,Nisini2015}). The comparison of detailed models with observations, therefore, allows one to address complex and unsolved issues such as the role of stellar and active galactic nuclei (AGN) feedback on the cycle of matter \citep{ostriker_maximally_2011} and galaxy evolution \citep{schaye_2015,hopkins_2018, richings_origin_2018} as well as the transfer of mass, momentum, and energy from the large scales to the viscous scale in turbulent multiphase media.

New examples of the ubiquity and modus operandi of interstellar shocks have emerged from  recent observations of extragalactic environments. In the Stefan's Quintet collision, for instance, a large scale $\sim 1000$ km/s shock produces X-ray emission from the $T > 10^6$ K shock-heated gas. Warm molecular hydrogen, which should be destroyed at these temperatures, dominates the cooling in this region \citep{guillard_2009}, revealing the multiphase nature of the environment along with the necessary transfer of kinetic energy to lower velocity shocks. Interestingly, CO and [CII] lines have linewidths as large as $1000$ km/s, bearing the signature of the turbulent cascade driven in the post-shock gas by the large-scale shock \citep{appleton_powerful_2017}. Similarly, shocked, warm molecular hydrogen emission has also been observed in a sample of 22 radio galaxies in which star formation is quenched \citep{Lanz2016}. The authors suggest that shocks driven by the radio jets inject turbulence into the interstellar medium, which powers the luminous  H$_2$ line emission. Further opening occurred with the discovery of ubiquitous, powerful molecular outflows in local and high-redshift starburst galaxies \citep{Veilleux2020,Hodge2020}. Molecular lines in galactic outflows are found with velocity dispersions in excess of $1000$ km/s, which can be seen as the signature of powerful turbulence generated by galactic winds \citep{falgarone_2017}. All of these examples reinforce the need for sophisticated and publicly accessible models capable of following the complex chemical and thermal evolution of shocks, down to the viscous scale, in a variety of physical conditions and for different input of mechanical energy.

Detailed shock models have long been developed for a variety of applications in the interstellar medium. Many works have been dedicated to modelling the heating, ionisation and radiative signatures of high velocity shocks (typically $V_s> 100$~km/s) propagating into the atomic environments (with ambient gas densities $ n<1$~cm$^{-3}$) found in the intercloud medium \citep{pikelner_energy_1957}, supernovae remnants \citep{cox_theoretical_1972, dopita_optical_1976, shull_theoretical_1979} and Herbig-Haro objects \citep{raymond_shock_1979}. Particular attention has been given to accurately treating the radiative precursor, the region ahead of the shock affected by photons generated by the shock itself. Photoionisation in this region can alter the state of the gas before it enters the shock front and thus determines the shock initial conditions. Most recently, state of the art models draw from increasingly large atomic databases to accurately predict the spectra of radiative atomic shocks \citep{hartigan_new_2015,sutherland_effects_2017,dopita_effects_2017}.

There have also been extensive studies of shocks focusing on molecular environments. Early work by \cite{field_hydromagnetic_1968} and \cite{aannestad_molecule_1973} considered molecular chemistry in low velocity shocks ($V_s < 20$~km/s) in dense environments ($n=10$--$100$~cm$^{-3}$), with a focus on the survival of molecules through the shock-heated gas. These shocks are not hot enough to produce significant UV photons, and hence no treatment of a radiative precursor is required. In a series of works, \cite{hollenbach_molecule_1979,hollenbach_molecule_1980,hollenbach_molecule_1989} treated the reformation of molecules in the cooling flow behind a fully dissociated shocked gas. By using the UV field calculated by a specialised atomic shock code, they were able to make an extensive study of shocks over both low and high velocities ($V_s=25-150$~km/s) and a broad range of densities ($n=10$--$10^9$~cm$^{-3}$). This work thoroughly detailed the key physical processes---such as photoionisation and dissociation, formation of H$_2$ on dust, cooling from both atoms and molecules---which determine the structure and radiative characteristics of shocks in these environments. Further advancements were made by \cite{neufeld_fast_1989} focusing on an accurate treatment of the UV radiative transfer, particularly Ly$\alpha$ and two-photon continuum, and its impact on molecular chemistry in intermediate velocity shocks ($V_s=60$--100~km/s) in dense media ($n=10^4$--$10^6$~cm$^{-3}$). 

In the present paper, we revisit these pioneering and sophisticated developments by implementing the physics and chemistry of self-irradiated shocks in the Paris-Durham code\footnote{available on the ISM platform https://ism.obspm.fr}, a public and versatile state-of-the-art tool initially designed for the study of molecular shocks. Building on the recent updates of \citet{lesaffre_low-velocity_2013} and \citet{godard_models_2019} who focused on low velocity shocks ($V_s \leqslant 25$ km s$^{-1}$) irradiated by an external radiation field, we explore here an intermediate velocity range (25 km s$^{-1}$ $\leqslant V_s \leqslant 60$ km s$^{-1}$) where shocks are hot enough to generate significant UV radiation, yet cool enough to prevent the production of multi-ionised species. We present tracers for this velocity range, including atomic line emission, the Ly$\alpha$ and Ly$\beta$ counterparts, as well as tracers resulting from the complex molecular chemistry in the cooling flow. Infrared atomic and molecular lines are particularly timely given the wealth of data coming from the \textit{Atacama Large Millimeter/submillimeter Array} (ALMA) and in anticipation of the \textit{James Webb Space Telescope} (JWST).

The Paris-Durham shock code is introduced in section~\ref{sec:shock} where we describe the updates implemented for this work. As a first step to compute the self-generated UV field we only treat the emission from atomic hydrogen. The corresponding three-level atomic model is outlined in section~\ref{sec:atomH}. In section~\ref{sec:radtrans} we describe the post-processed radiative transfer treatment to compute the shock generated radiation field. The effects of this radiation field on shocks at different velocities are shown in section~\ref{sec:results}. We finally conclude by summarising our results in section~\ref{sec:conclusion}.

%
%
\section{Shock model}\label{sec:shock}
In order to self-consistently compute a shock and the UV radiation field it generates, we iteratively solve these two components separately. We first generate a shock solution with the Paris-Durham shock code, which we describe in this section, then post-process this output to compute the radiation field. We then run the shock code again in the presence of this radiation field and repeat until convergence.

\subsection{J-type shocks}

\begin{figure}
\centering
  \includegraphics[width=\columnwidth, trim=0.5cm 0cm 0cm 0cm ]{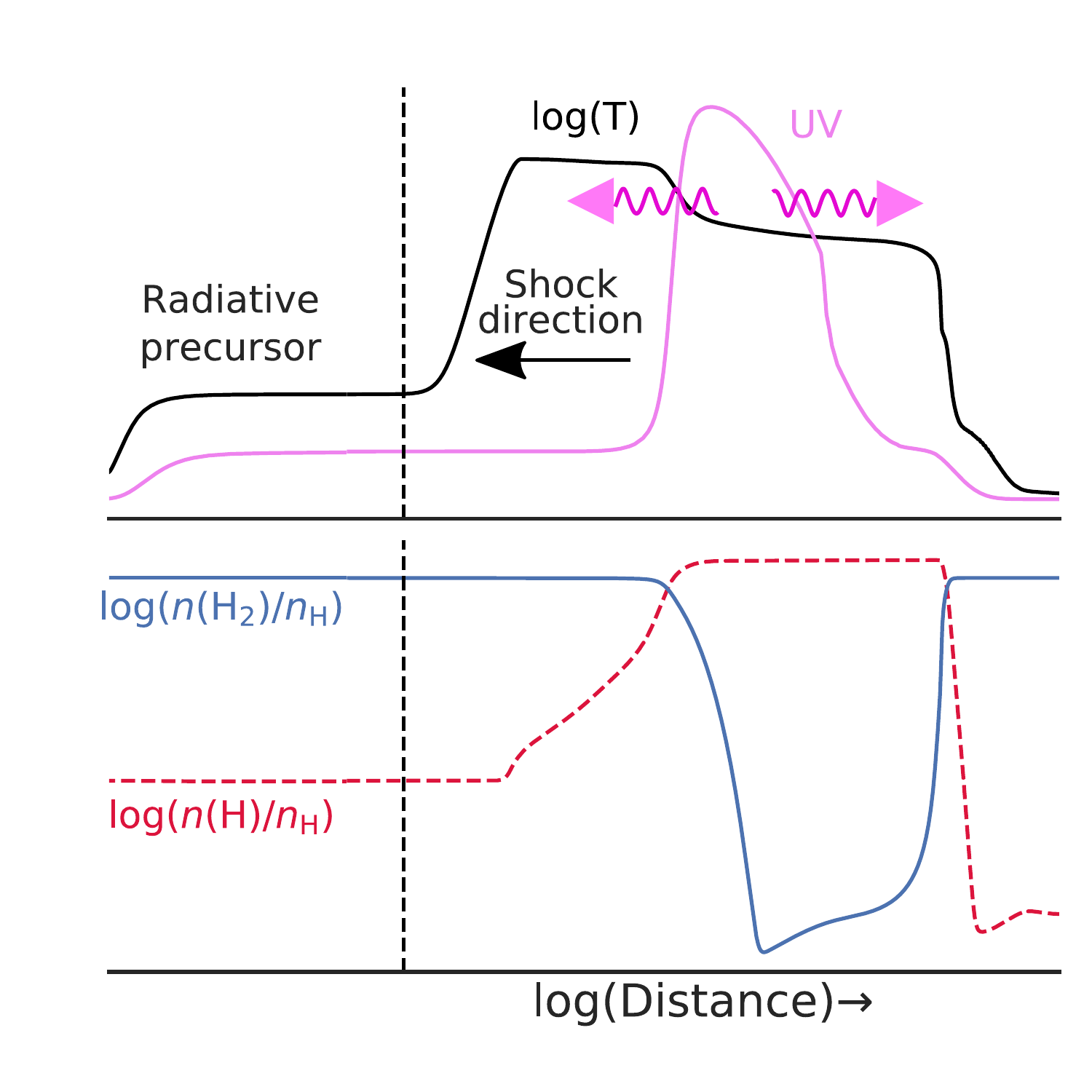}
  \caption{Scheme of the profiles of temperature and UV production (top), and hydrogen abundances (bottom), for typical J-type molecular shocks. The shocked region (right of the dashed vertical line) is shown in log-scale of the distance in order to emphasise the initial adiabatic jump.}
  \label{fig:profile}
\end{figure}

Interstellar shocks come in a variety of flavours, depending on the shock speed in relation to the signal speeds in the neutral and ionised components of the fluid \citep{draine_1993}. We focus solely on single-fluid J-type (discontinuous) shocks in this work, because C-type (continuous) shocks do not reach the high temperatures required to generate significant UV radiation. In addition, for the conditions we consider (Table~\ref{tab:Hparams}), in particular for weak magnetic fields, C-type shock velocities are bounded to velocities $V_s < 25$~km/s (see figure D1 in \cite{godard_models_2019}). Figure~\ref{fig:profile} shows the typical evolution of the temperature and atomic and molecular hydrogen abundances in a J-type molecular shock. Viscous heating mediates an initial adiabatic jump in temperature up to a first plateau, whose temperature can be estimated using the Rankine-Hugoniot jump conditions, assuming an adiabatic index of $5/3$, shock velocity much larger than Alfven velocity ($v_{\rm A}=B/\sqrt{4\pi\rho}$ for magnetic field $B$ and fluid mass density $\rho$) and fully molecular medium (mean molecular weight of 2.33 a.m.u.), as
\begin{align}\label{eq:rankineT}
T_{\rm{max}} \sim 53 \, \rm{K} \, \left( \frac{V_s}{\rm{km/s}} \right)^2
\end{align}
where $V_s$ is the shock speed. At velocities greater than $\sim$30~km/s interstellar shocks can reach temperatures sufficient to collisionally dissociate H$_2$, with dissociation temperature of 5.2$\times 10^4$~K, and to collisionally excite atomic hydrogen, with electronic levels starting at $\sim$1.2$\times 10^5$~K. Atomic H is therefore produced in the hottest parts of the shock. Strong cooling due to Ly$\alpha$ excitations causes the temperature to drop from the peak to settle at $T\lesssim$~10$^4$~K. This plateau is maintained until H$_2$ reforms on grains deeper in the shock. In the transition between the adiabatic and second temperature plateau the excitation of H is strong enough to produce significant amounts Ly$\alpha$ photons. \cite{flower_excitation_2010} estimated that a 30~km/s shock with preshock density $n_{\rm{H}}=2\times$10$^5$~cm$^{-3}$ would produce a Ly$\alpha$ photon flux three orders of magnitude more than that of the standard UV interstellar radiation field (ISRF, \cite{mathis_interstellar_1983}). Such an effect calls for a detailed treatment of the UV radiative transfer in molecular shocks, and hence as a first step in this direction we consider the addition of just one source of UV photons, the hydrogen atom. After summarising key aspects of the Paris-Durham shock code and outlining the updates used in this work we describe the 5-level model of hydrogen in section~\ref{sec:atomH}.

\subsection{Paris-Durham shock code}
The Paris-Durham shock code gives the steady-state solution of the plane-parallel magnetohydrodynamics equations coupled with cooling functions and an extensive chemical reaction network appropriate to the molecular phases of the interstellar medium. The version used in this work is that of \cite{flower_influence_2003}, with updates described in \cite{lesaffre_low-velocity_2013}, \cite{godard_models_2019} and here.

\begin{table}
\centering
\caption{Fractional gas-phase elemental abundances used in this work. We adopt the elemental abundances of \cite{flower_influence_2003} and put all of the species depleted on grain mantles into the gas phase. Numbers in parentheses denote powers of 10.}
\begin{tabular}{l c} \cline{1-2}\cline{1-2}
\multicolumn{2}{c}{\vspace{-0.3cm}} \\ 
Element & Abundance \\ \cline{1-2}
\multicolumn{2}{c}{\vspace{-0.3cm}} \\ 
H  & 1.00     \\
He & 1.00(-1) \\
C  & 1.38(-4) \\
N  & 7.94(-5) \\
O  & 3.02(-4) \\
Si & 3.00(-6) \\
S  & 1.86(-5) \\
Fe & 1.50(-8) \\ \cline{1-2}
\multicolumn{2}{c}{\vspace{-0.3cm}}
\end{tabular} \label{tab:elements}
\end{table}

The code solves a set of coupled, first-order, ordinary differential equations---given by \cite{flower_flows_2010}---using the DVODE forward integration algorithm \citep{hindmarsh_1983} from some initial conditions. The dynamical variables, that is to say the density and velocity, are often parameters with initial values that we choose to explore, but the initial temperature and chemical abundances are calculated by solving the chemical and thermal equations in a uniform slab for $10^7$~years so that the material entering the shock is roughly in chemical and thermal equilibrium. We adopt the elemental abundances of \cite{flower_influence_2003} and put all of the species depleted on grain mantles into the gas phase (see Table~\ref{tab:elements}). In order to integrate through the discontinuity in J-type shocks, the code employs an artificial viscosity treatment \citep{richtmyer_1957}. This treatment has been verified to uphold the Rankine-Hugoniot relations, producing a smooth adiabatic jump in which momentum and energy are conserved.

In J-type shocks with molecular initial conditions the dominant source of atomic H in the hottest parts of the shock is the collisional dissociation of H$_2$. The treatment of this process in the code has been detailed in \cite{flower_structure_1996} and \cite{le_bourlot_2002}, but we summarise it here in order to stress its importance for the accuracy of Ly$\alpha$ production in molecular shocks. The code takes into account the reduced energy required to dissociate H$_2$ when excited in its rovibrational states. The dominant collision partners are H, H$_2$, H$^+$, and e$^-$. Excitations to the dissociative triplet state b$^3\Sigma_u^+$ by collisions with electrons has a rate coefficient given by
\begin{align}
k_e(T) &= 2\times 10^{-9}  \, {\rm cm}^3 {\rm s}^{-1} \left(\frac{T}{300~{\rm K}} \right)^{1/2} \times \nonumber \\
 & \sum_{v,J} f_{\rm{H}_2}\left(v,J\right) \exp \left(-\frac{E_T - E\left(v,J \right)}{T} \right)
\end{align}
where $E_T=116300$~K is the excitation energy of the triplet state and $E\left(v,J\right)$ and $f_{\rm{H}_2}\left(v,J\right)$ are the energy and fractional population of H$_2$ in the rovibrational state $v$, $J$, respectively. For collisions with H, the dissociation rate coefficient is given by
\begin{align}
k_{\rm H}(T) &= 10^{-10}  \, {\rm cm}^3 {\rm s}^{-1}  \sum_{v,J} f_{\rm{H}_2}\left(v,J\right) \exp \left(-\frac{E_D - E\left(v,J \right)}{T} \right)
\end{align}
where $E_D=52000$~K (4.48~eV) is the dissociation energy of H$_2$. Rate coefficients for collisions with H$_2$ are unknown for astrophysical conditions, but we use a rate 8 times lower than for H collisions, as indicated by shock tube experiments \citep{jacobs_kinetics_1967,breshears_1973}.

The time-dependent populations of the rovibrational levels are solved in parallel with the dynamical and chemical variables as described in \cite{le_bourlot_2002} and \cite{flower_contributions_2003}, allowing for collisional and radiative transitions between the levels. In Appendix~\ref{app:h2levels} we check the effect of the number of levels treated has on the shock structure. We found that treating 150 levels is both computationally feasible and accurately models the H$_2$ cooling and dissociation.

\subsection{Updates}
As we post-process the UV radiative transfer (section~\ref{sec:radtrans}), the main update to the shock code is that it reads a given UV field at any point during the shock integration. Here we describe the modifications to account for this local field in addition to other updates on previous versions.

\subsubsection{Photoreactions}
Our treatment of photo-reactions (dissociation and ionisation) is slightly modified from \cite{godard_models_2019}, because we don't have to account for the attenuation of the radiation as this is done in the post-processed radiative transfer. Given the local radiation field energy density---calculated in the post-processed radiative transfer described in section~\ref{sec:radtrans}---defined as
\begin{align}\label{eq:u_isrf}
u_\nu = \frac{4\pi}{c} J_\nu
\end{align}
where $c$ is the speed of light and $J_\nu$ is the angle averaged specific intensity at frequency $\nu$, the photo-reaction rate is given by
\begin{align}
k_\gamma = c\int \frac{u_\nu}{h\nu} \sigma_\nu d\nu
\end{align}
where $h$ is Planck's constant and $\sigma_\nu$ is the frequency dependent cross-section. We use the cross sections of \cite{tabone_molecule_2020} for the most important photoreactions, the photo-ionisation of C, S, and Si, and both photo-ionisation and -dissociation of CH, OH, H$_2$O, O$_2$, and C$_2$. For all other photoreactions, we use the form
\begin{align}
k_\gamma = G_{\rm{eff}} \alpha
\end{align}
where $\alpha$ is the rate in the unattenuated standard ISRF, and $G_{\rm{eff}}$ is the effective radiation parameter integrated over the whole frequency range (corresponding to wavelengths between 911--2400~\AA), defined as
\begin{align}\label{eq:geff}
G_{\rm{eff}} = \frac{\int d\nu \, c u_\nu / h\nu }{1.55\times 10^8 \, \rm{ph/s/cm}^2}.
\end{align}
We note that $G_{\rm{eff}}$ is simply the flux of photons normalised to the \cite{mathis_interstellar_1983} ISRF with modifications described in \cite{tabone_molecule_2020}.

\subsubsection{Atomic H cooling} \label{sec:Hcooling}
The energy lost due to the collisional excitation of the electronic levels of H is given by
\begin{align}
\mathcal{B} = \sum \left(C_{ij} n_i - C_{ji} n_j \right) \Delta E_{ij}
\end{align}
where the $C_{ij}$ and $\Delta E_{ij}$ are, respectively, the collision rates and energy differences between levels $i$ and $j$, given in the next section, and the summation is taken over all pairs of levels. During the first shock iteration we compute the level populations $n_i$ by solving for statistical equilibrium treating only collisions and spontaneous emission. With no absorption or stimulated emission, this is an optically thin treatment. In subsequent iterations of the shock solution, we skip this computation and instead use the populations found during the post-processed radiative transfer described in section~\ref{sec:radtrans}. In this way the shock solution has a self-consistent treatment of cooling due to H.

\subsection{Chemical network}
Starting with the reaction network of \cite{godard_models_2019}, we remove all adsorption and desorption reactions as we are interested in shocks fast enough to produce a UV field that removes all mantles via photodesorption. We then extend the network to include collisional ionisation and dissociation rates from \cite{hollenbach_molecule_1989}. These include ionisation of He, all metals and select molecules---OH, H$_2$O, O$_2$, CH, and CO---by collisions with H, H$^+$, He, and electrons. We also include collisional dissociation by electrons for H$_2$O, OH, and CO. These updates result in a chemical reaction network including 1256 reactions for 141 species.

%

\section{Atomic hydrogen parameters}\label{sec:atomH}
\subsection{Collisional rates}

\begin{figure}
\centering
  \includegraphics[width=\columnwidth]{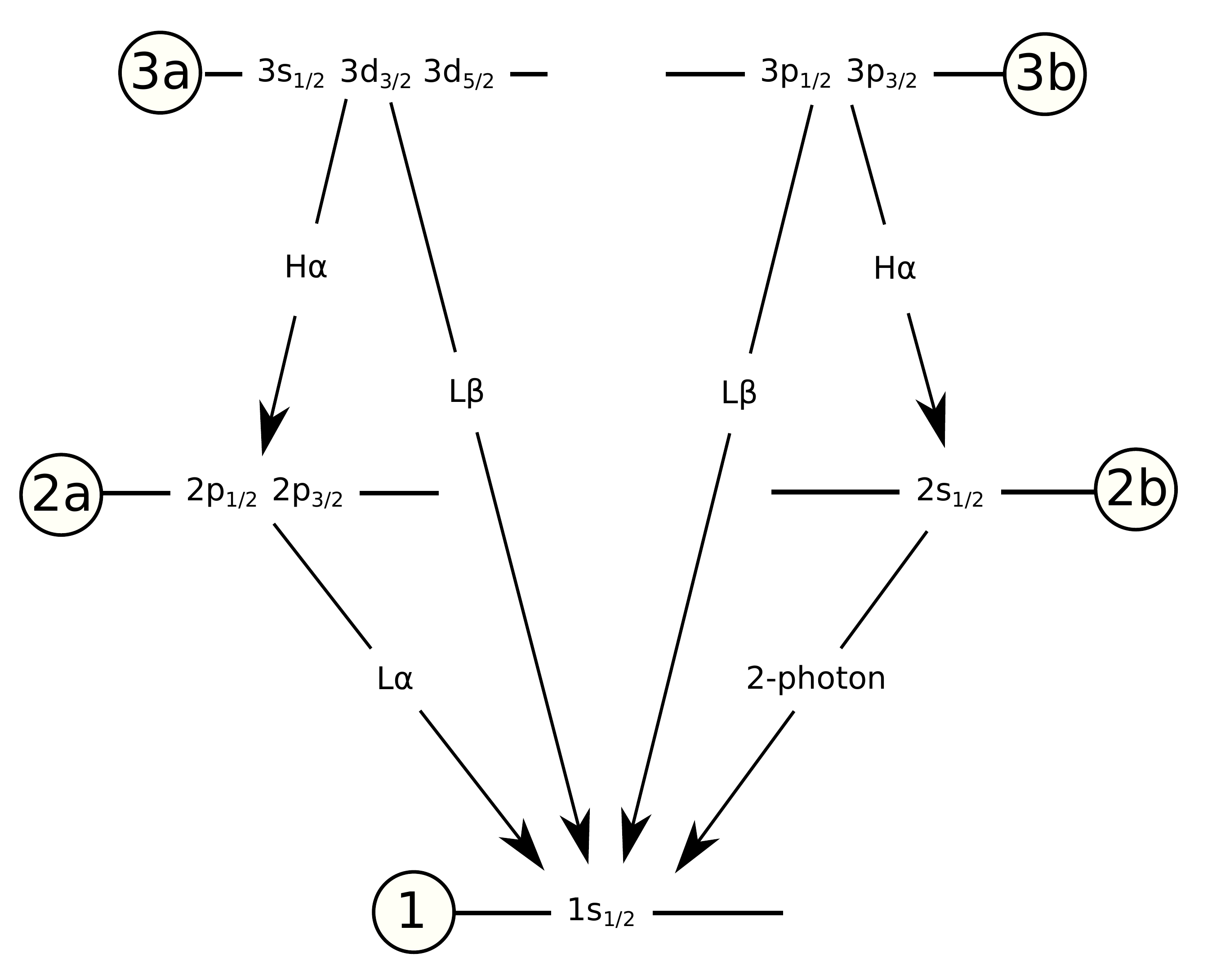}
  \caption{Three level atomic hydrogen model and its radiative transitions. Statistical weights are $g_1=2$, $g_{2a}=6$, $g_{2b}=2$, $g_{3a}=12$ and $g_{3b}=6$.}
  \label{fig:5lvl-hydro}
\end{figure}

 

\begin{table*}
\centering
\caption{Atomic hydrogen radiative and collisional transition parameters.}
\begin{tabular}{l l c c c c c r} \cline{1-8}\cline{1-8}
\multicolumn{8}{c}{\vspace{-0.3cm}} \\ 
\# & Transition & $\lambda \, (\AA)$ & $A_{ij}$ (s$^{-1}$) & $\Delta E$~(K) & Col. Partner &  $\alpha$ (cm$^3$ s$^{-1}$) & $\beta$\\ \cline{1-8}
\multicolumn{8}{c}{\vspace{-0.3cm}} \\ 
1  & 2a$\to$1 (Ly$\alpha$) & 1215.67 & $6.265\times 10^8$ & 118352.595 & e$^{-}$ & $4.329\times 10^{-9}$ & $0.0984$  \\
2  & 2b$\to$1 (2ph)       & 1215.67 & $8.200$            & 118352.295 & e$^{-}$ & $4.195\times 10^{-8}$ & $-0.3591$ \\
3  & 3a$\to$1 (Ly$\beta$)  & 1025.72 & $4.948\times 10^2$ & 140269.714 & e$^{-}$ & $3.075\times 10^{-9}$ & $-0.3358$ \\
4  & 3b$\to$1 (Ly$\beta$)  & 1025.72 & $1.673\times 10^8$ & 140269.659 & e$^{-}$ & $1.641\times 10^{-9}$ & $-0.0132$ \\
5  & 3a$\to$2a (H$\alpha$) & 6564.63 & $3.662\times 10^7$ & 21917.118  & e$^{-}$ & $3.331\times 10^{-8}$ & $0.0186$  \\
6  & 3b$\to$2b (H$\alpha$) & 6564.63 & $2.245\times 10^7$ & 21917.364  & e$^{-}$ & $1.590\times 10^{-7}$ & $0.1537$  \\ \cline{1-8}\multicolumn{8}{c}{\vspace{-0.3cm}} \\ 
7  & 3a$\to$2b             &         &                    & 21917.418  & e$^{-}$ & $9.478\times 10^{-8}$ & $-0.1585$ \\
8  & 3b$\to$2a             &         &                    & 21917.064  & e$^{-}$ & $1.614\times 10^{-8}$ & $0.2123$  \\
9  & 2b$\to$2a             &         &                    & -0.300     & e$^{-}$ & $2.112\times 10^{-4}$ & $-0.3735$ \\ 
10 & 2b$\to$2a             &         &                    & -0.300     & H$^{+}$ & $8.039\times 10^{-4}$ & $-0.1507$ \\\cline{1-8}
\multicolumn{8}{c}{\vspace{-0.3cm}}
\end{tabular}
  
\emph{Notes}. Radiative transitions 7--10 are forbidden and are not taken into account.
\label{tab:Hparams}
\end{table*}

In order to study the UV field generated inside a molecular shock, we consider a three-level atomic hydrogen model including the Ly$\alpha$ ($n=2\to 1$), Ly$\beta$ ($n=3\to 1$), and H$\alpha$ ($n=3\to 2$) line transitions. The second level is divided into 2 sub-levels in order to model the two-photon emission from the $2s$ metastable state. The third level is also divided into two, grouped by orbitals with radiative transitions to either the $2p$ or $2s$ orbitals. This division of sub-levels is schematically shown in Fig.~\ref{fig:5lvl-hydro} and the values of the transition wavelengths, Einstein $A$ coefficients, and energy differences $\Delta E$ for the 6 radiative transitions use atomic data from the NIST database \citep{NIST_ASD} and are listed in Table~\ref{tab:Hparams}.

Collisions with electrons dominate most of the rates so we consider mostly H-e$^-$ collisions. The collisional de-excitation rates coefficients are given by
\begin{align}
k_{ij} &= \frac{h^2}{2 \pi m_e^2}\left(\frac{m_e}{2 \pi k_B T}\right)^{1/2}\frac{\Upsilon_{ij}}{g_i}
\end{align}
where $\Upsilon_{ij}$ is the effective collision strength. 
We fit $\Upsilon_{ij}$ data of \cite{anderson_2002} (given in Appendix~\ref{app:hydrogen}) as a temperature dependent power-law so that the de-excitation rate can be written as
\begin{align}\label{eq:deex}
k_{ij} &= \alpha \left(\frac{T}{300 \, \rm{K}} \right)^\beta \,\, \rm{cm^3 \, s^{-1}}.
\end{align}
Collisional rate coefficients for the $2s$ to $2p$ orbitals are stronger for H-H$^+$ collisions than H-e$^-$ collisions, so we also include these rates as given by \cite{osterbrock_2006} fit by the same power law equation~\eqref{eq:deex}. The collision rate parametres $\alpha$, $\beta$, and energy differences $\Delta E$ for the 10 collisional transitions are listed in Table~\ref{tab:Hparams}.

\subsection{Non local thermodynamic equilibrium parameter}
A first characterisation of a radiative transfer problem is given by the non local thermodynamic equilibrium (LTE) parameter
\begin{align}
\epsilon = \frac{ C_{ij} } { C_{ij} + A_{ij}/\left(1 - \exp\left(-\Delta E_{ij}/k_B T\right) \right)}
\end{align}
for a line with collisional de-excitation rate $C_{ij}=k_{ij}n_C$ where $n_C$ is the number density of the collisional partner. $\epsilon$ characterises the competition between the re-emission of an absorbed photon versus its destruction due to collisional de-excitation, and so approximates a photon destruction probability. If $\epsilon=1$, collisions dominate so that LTE holds and the radiation is thermal. As $\epsilon \to 0$ the radiative transfer becomes increasingly difficult as the effects of scattering become important. In the fiducial shock we consider in section~\ref{sec:fiducial} the Ly$\alpha$ emission is generated in gas with temperature $T\sim 10^4$~K and electron density $n_e\sim 10^3$~cm$^{-3}$, giving a non-LTE parameter $\epsilon \sim 10^{-14}$. This parameter appears in the solution of the statistical equilibrium equations and at such a low value can cause important rounding errors with double precision variables. We thus go to quad precision to avoid this problem. In the next section we turn to the Accelerated Lambda Iteration method to overcome numerical difficulties caused by extreme optical depths.

%
%
\section{Post-processed radiative transfer}\label{sec:radtrans}
In this section we describe the post-processing of the Paris-Durham outputs to obtain the UV field at every point inside the shock as well as extending into the unshocked region ahead of the shock. This region ahead of the shock influenced by these UV photons is called the \textit{radiative precursor}.
  
To briefly summarise, we first use the Accelerated Lambda Iteration (ALI) algorithm rendered necessary by the extreme optical  depths in the Ly$\alpha$ transition. We then use the ALI calculated excited level populations to solve the radiative transfer for the two-photon continuum emission from the $2s$ metastable state. Finally the radiation field is extended into the radiative precursor. We give a summary of the ALI algorithm here, but for a detailed review see \cite{hubeny_accelerated_1992} and references therein.

\subsection{Accelerated Lambda Iteration}
We seek to calculate the angle-averaged UV intensity
\begin{align}\label{eq:mean_intensity2}
J_\nu &= \frac{1}{4\pi}\int I_{\mu\nu} d\Omega
\end{align}
at each position, $z$, in the shock in order to compute the UV energy density (equation~\ref{eq:u_isrf}) that is used to calculate photoelectric heating, photo-ionisation, and dissociation rates in the Paris-Durham shock code. For a plane-parallel semi-infinite slab, the specific intensity on a ray, $I_{\mu\nu}$, with angle $\mu=\cos\theta$ with respect to the slab normal satisfies the Radiative Transfer Equation (RTE)
\begin{align}
\mu \frac{d I_{\mu\nu}}{dz} = j_\nu - \kappa_\nu I_{\mu\nu}
\end{align}
where the line emission coefficient for the transition $i \to j$ 
\begin{align}
j_\nu = \frac{h \nu_{ij}}{4\pi} A_{ij} n_i \phi_{\mu\nu}
\end{align}
and line opacity
\begin{align}
\kappa_\nu &= \frac{h \nu_{ij}}{4\pi} \left(B_{ji} n_j - B_{ij}n_i \right) \phi_{\mu\nu} 
\end{align}
where $n_i$ and $n_j$ are the densities of the upper and lower levels of atomic hydrogen respectively, and $B_{ij}$ and $B_{ji}$ are the Einstein $B$ coefficients. The Gaussian line profile is given by
\begin{align}\label{eq:lineprofile}
\phi_{\mu\nu} = \frac{c}{v_T \nu_{ij} \sqrt{2\pi}} \exp \left( -\frac{1}{2}\left( \frac{\Delta \nu}{\nu_{ij}} \frac{c}{v_T} \right)^2 \right)
\end{align}
where the thermal velocity
\begin{align}
v_T = \sqrt{\frac{k_B T}{m_H}}
\end{align}
and frequency shift from the Doppler-shifted line centre, $\nu_{ij}$, 
\begin{align}
\Delta \nu = \nu - \nu_{ij}\left(1 + \frac{v_z \mu}{c} \right)
\end{align}
where $v_z$ is the flow velocity in the shock propagation direction.
  
The level populations are needed to compute the emission coefficients, and are obtained by solving the Statistical Equilibrium (SE) equations
\begin{align}\label{eq:statequil}
\dot{n}_i = 0 = & \sum_{j\neq i} \left(C_{ji} n_j - C_{ij} n_i\right) + \sum_{j > i} \left( A_{ji} n_j + \left(B_{ji}n_j - B_{ij}n_i \right)\bar{J}_{ij} \right) - \nonumber \\ 
& \sum_{j < i} \left( A_{ij} n_i + \left(B_{ij}n_i - B_{ji}n_j \right)\bar{J}_{ij} \right)
\end{align}
where the $\bar{J}_{ij}$ are the mean intensities, $ J_{\nu}$, averaged over the line profile for that transition. In order to directly solve this at each position in the shock one would need to invert an enormous matrix with dimensions determined by the discretisation choices for the number of grid positions, frequencies, and angles. As this is computationally unrealistic, the usual strategy is to iterate between solving the RTE and SE equations. Our algorithm is summarised as follows:
    
\begin{enumerate}
  
\item  We initialise the populations by solving the SE equations with the $\bar{J}_{ij}$ set to zero. The populations are used to compute the source function
\begin{align}
S_{ij} = \frac{j_\nu}{\kappa_\nu} = \frac{A_{ij}n_i}{B_{ji} n_j - B_{ij} n_i}
\end{align}
where we have assumed complete frequency redistribution, that is that the emission and absorption profiles are equal.

\item The source function is used to solve the RTE along some ray, giving the specific intensity at each point in the shock for every desired frequency. The solution can be written in terms of an integral operator, $\Lambda_{\mu\nu}$, acting on the source function
\begin{align}\label{eq:specific_intensity}
I_{\mu\nu} = \Lambda_{\mu\nu} S_{ij}.
\end{align}
On a discrete grid this can be written
\begin{align}\label{eq:partial_lambda}
I_{\mu\nu}\left( z_l \right) = \sum_{l'} \Lambda_{\mu\nu}\left(z_l,z_{l'}\right) S_{ij}\left(z_{l'} \right)
\end{align}
emphasising that the intensity at any one point is coupled to the source function at all points, and that this coupling is encoded in $\Lambda_{\mu\nu}$.
  
\item The intensity is used to compute the profile-integrated angle-averaged intensity for each transition
\begin{align}\label{eq:mean_intensity1}
\bar{J}_{ij} &= \frac{1}{2}\int_{-1}^1 d\mu \int I_{\mu\nu} \phi_{\mu\nu} d\nu.
\end{align}
We compute the intensity over two rays with angles $\theta=0$ and 180$^\circ$, which has two advantages. Firstly it is coherent with the energetics of the plane-parallel shock in which the energy flux changes only along this direction. Secondly it avoids the non-convergence of the integral caused by emission adding up over infinite length rays near $\theta=90^\circ$. 
  
\item The SE equations are solved to update the level populations. Without modification if we return to the first step and iterate to convergence we have the classical Lambda Iteration method, because the mean intensity in the SE equations is replaced by combining Eqs.~\eqref{eq:specific_intensity} and \eqref{eq:mean_intensity1}
\begin{align}
\bar{J}_{ij} &= \left( \frac{1}{2}\int_{-1}^1 d\mu  \int \Lambda_{\mu\nu} \phi_{\mu\nu} d\nu \right) S_{ij}\\
& = \Lambda_{ij} S_{ij}
\end{align}
where $\Lambda_{ij}$---the profile-integrated angle-averaged $\Lambda_{\mu\nu}$---is the $\Lambda$-operator for this transition. This iteration scheme is known to be pseudo-convergent in systems with extreme optical depths. This effect is shown in Appendix~\ref{app:twostream} where we also verify the method against analytic solutions. The ALI algorithm overcomes such pseudo-convergence by splitting the $\Lambda_{ij}$ operator
\begin{align}
\Lambda_{ij} &= \Lambda^*_{ij} + \Lambda_{ij} - \Lambda^*_{ij}
\end{align}
where $\Lambda^*_{ij}$ is the approximate $\Lambda$-operator. The iterative scheme is chosen such that at iteration $n$
\begin{align}
\left(\Lambda_{ij} S_{ij}\right)^n &=  \Lambda^*_{ij} S_{ij}^n + \left( \Lambda_{ij} - \Lambda^*_{ij}\right) S_{ij}^{n-1} \label{eq:split1}\\
&=  \Lambda^*_{ij} S_{ij}^n + \bar{J}_{ij}^{n-1}- \Lambda^*_{ij} S_{ij}^{n-1}.
\end{align}
\cite{olson_rapidly_1986} demonstrated mathematically that a nearly optimal choice of $\Lambda^*_{ij}$ is the diagonal of $\Lambda_{ij}$. By inspecting Eq.~\eqref{eq:partial_lambda}, the diagonal encodes the local contribution of the source function to the intensity at a given point. The splitting in Eq.~\eqref{eq:split1} can therefore be interpreted as solving the intensity exactly for the propagation within adjacent grid points while using the long range contribution from the previous iteration. 
  
With this choice of $\Lambda^*_{ij}=\rm{diag}\left(\Lambda_{ij}\right)$ the radiative terms in the SE equations become
\begin{align}
A_{ji} n_j(1 - \Lambda^*_{ij}) + \left(B_{ji}n_j - B_{ij}n_i \right)\left( \bar{J}_{ij}^{n-1} - \Lambda^*S_{ij}^{n-1}\right). 
\end{align}
Solving the SE equations with this modification, iteratively with Eq.~\eqref{eq:mean_intensity1} is the ALI algorithm.
  
\item Before going back to step 1 we further increase the rate of convergence of the iterative scheme by way of the Ng acceleration algorithm \citep{ng_hypernetted_1974} as formulated by \cite{olson_rapidly_1986}. This algorithm extrapolates the excited populations based on the previous three iterations. We only take this step if the column densities of all levels vary monotonically over those previous iterations.
  
\end{enumerate}
  
The approximate $\Lambda$-operator, $\Lambda^*_{ij}$, as the diagonal of $\Lambda_{ij}$ emerges from the calculation of the intensities. In step~2, to solve the RTE along a given ray we use the method of short characteristics formulated by \cite{paletou_fast_2007} in which the intensity at position $z$ is computed from the intensity at the upstream grid position, $z_u$, and the source function as
\begin{align}
I_{\mu\nu}(z) = I_{\mu\nu}(z_u) \exp \left( -\Delta \tau_u \right) + \int _0^{\Delta\tau_u} S(\tau)d\tau \label{eq:intensity_sol}
\end{align}
where the angle and frequency dependent upstream incremental optical depth is
\begin{align}
\Delta \tau_u = \int_{z_u}^z \kappa_\nu \left(z'\right) \frac{dz'}{\mu}.
\end{align}
The integral in Eq.~\eqref{eq:intensity_sol} is solved by assuming the source function is quadratic in the interval, interpolating on the grid points upstream ($z_u$) and downstream ($z_d$) of the point in question
\begin{align}
\int _0^{\Delta\tau_u} S(\tau)d\tau = \Psi_u S(z_u) + \Psi_o S(z) + \Psi_d S(z_d)
\end{align}
where the coefficients
\begin{align}
\Psi_u &= \frac{w_2 + w_1 \Delta \tau_d}{\Delta\tau _u\left(\Delta \tau_u + \Delta \tau_d\right)}\\
\Psi_o &= w_0 + \frac{w_1\left(\Delta \tau_u - \Delta \tau_d\right)}{\Delta \tau_u \Delta \tau_d} \label{eq:lambdastar}\\
\Psi_d &= \frac{w_2 - w_1 \Delta \tau_u}{\Delta\tau_d\left(\Delta \tau_u + \Delta \tau_d\right)}
\end{align}
where
\begin{align}
\Delta \tau_d = \int_z^{z_d} \kappa_\nu \left(z'\right) \frac{dz'}{\mu}
\end{align}
and
\begin{align}
w_0 &= 1 - \exp \left(-\Delta \tau_u\right) \\
w_1 &= w_0 - \Delta \tau_u \exp \left(-\Delta \tau_u\right) \\
w_2 &= 2 w_1 - \left(\Delta \tau_u \right)^2 \exp \left(-\Delta \tau_u\right).
\end{align}
The diagonal of the $\Lambda$-operator is then given by $\Psi_o$ integrated over all angles and frequencies weighted by the line profile. In Appendix~\ref{app:threelevel} we verify that the algorithm reproduces previous work on multi-level radiative transfer in idealised slabs.

\subsection{Two-photon emission}
The $2s$ metastable state of hydrogen cannot decay by a single photon process, and instead the $2s$-$1s$ transition proceeds by the emission of two photons. During the ALI computation this transition is treated as a resonance line with an Einstein coefficient $A_{2ph}=8.2$~s$^{-1}$ in order to determine the population of the $2s$ orbital. This population is then used to compute the continuum emission by solving the RTE including only dust opacity and an emission coefficient
\begin{align}
j_\nu = \frac{h \nu_{2ph}}{4\pi} A_{2ph} n_{2s} \psi_\nu
\end{align}
where $\psi_\nu$ is the functional form of the spectrum
\begin{align}
\psi_\nu = \frac{6\nu}{\nu_{2ph}^2} \left(1 - \frac{\nu}{\nu_{2ph}}\right)
\end{align}
used in \cite{shull_theoretical_1979}.

\subsection{Radiative precursor and postshock extension}
After computing the radiation field within the shock with the ALI method, we extend the radiation into the preshock by solving the RTE with the intensity at the shock front as a boundary condition. The ALI algorithm is not necessary in this region because of the low temperatures. In the first iteration, this escaping intensity impinges on a homogeneous slab entering the shock with flow velocity equal to the shock velocity. In subsequent iterations we use values of temperature and chemical composition computed with the Paris-Durham shock code.

To decide where to start the Paris-Durham integration the preshock is extended over a size, $L_{\rm{pre}}$, such that the dust attenuates the field that escapes the shock, $G_{\rm{eff},0}$, down to a negligible level compared to the ISRF. To do this we solve
\begin{align}
G_{\rm{eff}} = G_{\rm{eff},0} \exp \left( - \kappa_D L_{\rm{pre}}\right) = \frac{1}{3}
\end{align}
where $\kappa_D$ is the dust opacity at the Ly$\alpha$ central wavelength.

We similarly extend the radiation further into the postshock by solving the RTE with the ALI output intensity as a boundary condition. This boundary is chosen at a location in the shock where the line radiative transfer has forced the Ly$\alpha$ energy into the wings but before dust attenuation takes effect. 

In solving the RTE for the pre and postshock regions we include the opacity due to H, H$_2$, and dust. While many H$_2$ lines of the Lyman (B$^1 \Sigma^+_u$ -- X$^1 \Sigma^+_g$) and Werner (C$^1 \Pi_u$ -- X$^1 \Sigma^+_g$) band systems overlap with the Ly$\alpha$ or Ly$\beta$ emission, in practice we found only the $v=6-0$~P(1) line to be of importance. In the first iteration, the H$_2$ level populations are assumed thermal, and then we use the output of the shock code for self-consistent populations in subsequent iterations.

%
%
\section{Results}\label{sec:results}

\begin{figure*}
\centering
  \includegraphics[width=0.9\textwidth, trim=3cm 0cm 4cm 0cm ]{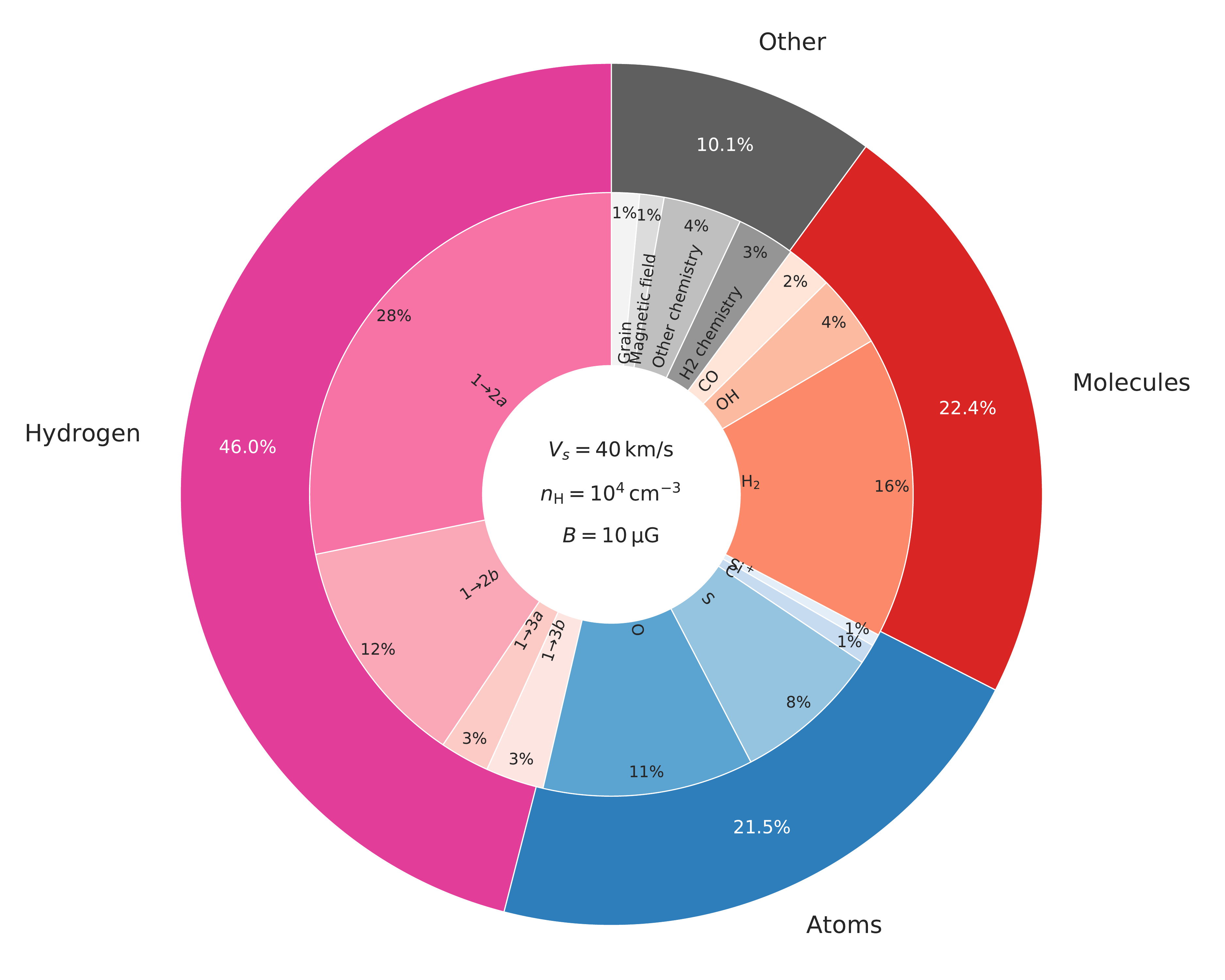}
  \caption{Pathways of energy reprocessing in the fiducial shock (Table~\ref{tab:shockparams}). This shows the energy lost due to excitation of atomic H, other atoms, molecules, and other processes as a percentage of the total energy flux. H$_2$ chemistry involves cooling due to collisional dissociation and heating due to reformation. Other chemistry is mostly cooling due to collisional ionisation.}
  \label{fig:rosace}
\end{figure*}

\begin{table}
\caption{Fiducial shock parameters.}
{\centering
\begin{tabular}{l c c} \cline{1-3}
\multicolumn{3}{c}{\vspace{-0.3cm}} \\
Parameter & Symbol & Value \\ \cline{1-3}
\multicolumn{3}{c}{\vspace{-0.3cm}} \\
Shock velocity & $V_s$ & 40 km/s \\
Proton density & $n_{\rm{H}}$  & $10^4$~cm$^{-3}$ \\
Magnetic field strength & $B$ & 10~${\rm \mu G}$ \\
Cosmic-ray ionisation rate & $\zeta$ & 5$\times$10$^{-17}$~s$^{-1}$ \\
External radiation field & $G_0$ & 0 \\
Viscous length & l & $10^9$~cm \\
H$_2$ levels & N$_{\rm{lev}}$ & 150 \\ \cline{1-3}
\end{tabular}} \label{tab:shockparams}
\end{table}
  
To understand the impact of the new treatment of the self-generated UV field we first consider a single fiducial case of a typical shock. We highlight the impact of this treatment by comparing to a shock model run without self-generated UV. After considering the effects on one shock, we compute shocks at a range of velocities to analyse trends with velocity and give observable predictions for molecular shocks at intermediate velocities.
  
\subsection{Fiducial case}\label{sec:fiducial}
As a fiducial case we consider a 40~km/s shock propagating into gas with total hydrogen density $n_{\rm{H}}=10^4$~cm$^{-3}$ and magnetic field strength 10~$\mu$G. In order to emphasise the adiabatic plateau, we adopt here an artificial viscous length of $10^9$~cm, 2 orders of magnitude smaller than the real viscous length deduced from the H$_2$-H$_2$ collision cross section. We note, however that the results are weakly dependent on this choice as long as the chosen viscous length is smaller than the real viscous length. In Table~\ref{tab:shockparams} we list the shock parameters. The shock acts to reprocess energy through heat into various atomic and molecular lines, dust emission, and to overcome energy barriers in chemical reactions. Some of this energy is also converted into magnetic energy. For the fiducial shock, all the components of this reprocessed energy are shown in Fig.~\ref{fig:rosace}. Excitation of molecules and atoms results in emission which escapes the shock and can be used to probe the shock properties. Excitation of H---though it is the largest component reprocessing almost half of the total energy flux---results in the emission of Ly$\alpha$, Ly$\beta$, and two-photon continuum UV radiation that is eventually absorbed by dust. Roughly half of this emission escapes ahead of the shock and is absorbed over a large distance set by the lengthscale of dust attenuation. The other half is reprocessed, in the cooling flow of the shock, into thermal energy via the photoelectric effect. Interestingly, the local cooling rate induced by collisional excitations of H and the subsequent interactions of Ly$\alpha$, Ly$\beta$, and two-photon continuum emission with the surrounding gas are not very different from the results obtained with an optically thin treatment of Ly$\alpha$ and Ly$\beta$. The convoluted radiative transfer presented in the previous section is important only to determine the exact line profile and the spatial asymmetry of the H emission, that is the exact fractions of the radiative energy that travel ahead of the shock and in the postshock.

\begin{figure*}[h!]
\centering
\includegraphics[width=\textwidth]{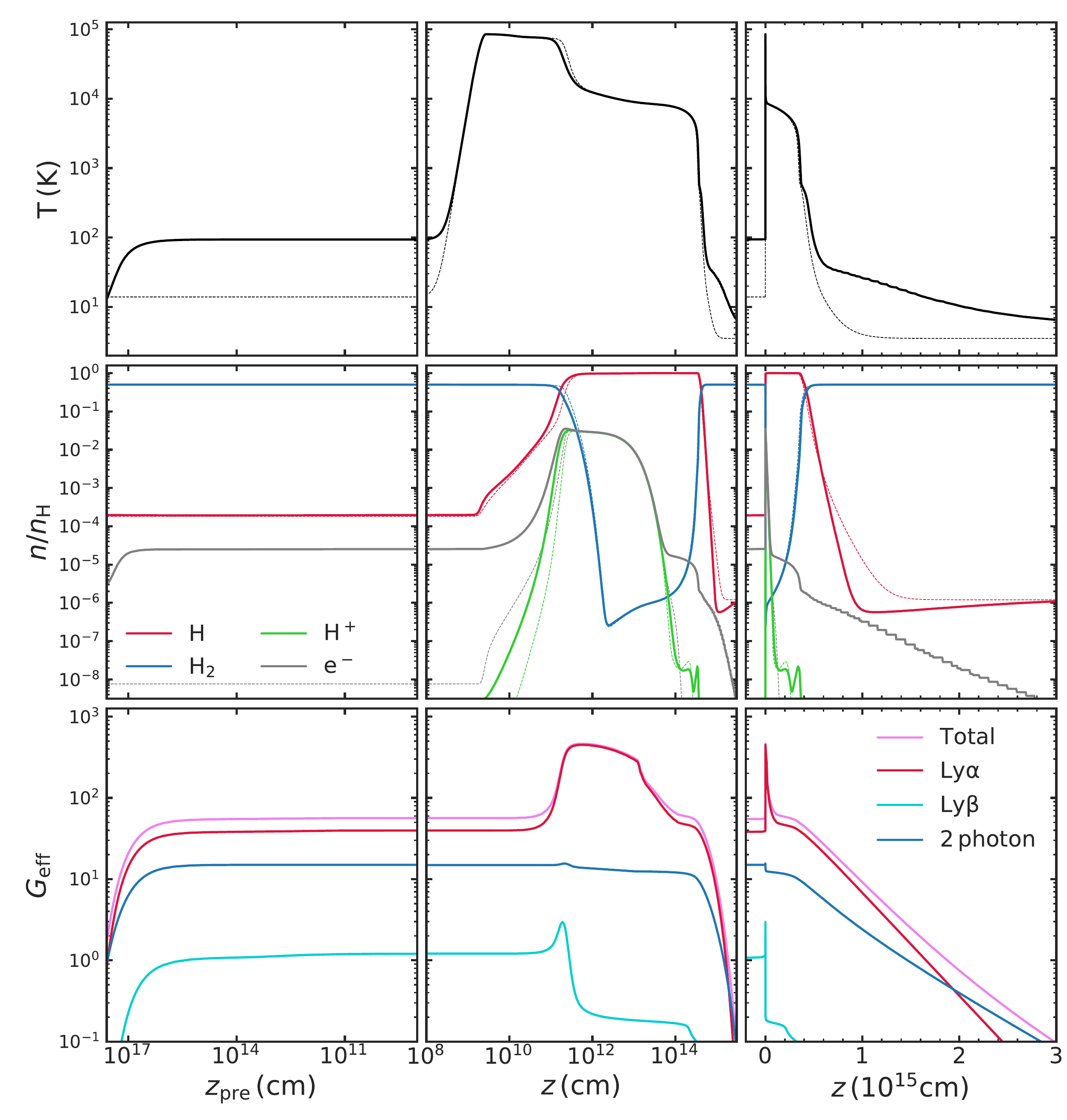}
\caption{Profiles of temperature (top), atomic, molecular, and ionised hydrogen densities (middle), and radiation field (bottom) for the fiducial shock (Table~\ref{tab:shockparams}). For the radiation field, we show the total emission computed with equation~\ref{eq:geff} (violet) and the contributions to this total by Ly$\alpha$ (red), Ly$\beta$ (cyan), and two-photon continuum (blue). The left column shows quantities in the radiative precursor, with distance increasing towards the left (i.e. distance from the shock front), the middle column shows postshock quantities with distance increasing towards the right in log scale while the right column is the same as the middle but in linear scale. Thick lines show results including the UV treatment, while thin dashed lines have no UV included.}\label{fig:profile_40}
\end{figure*}

\begin{figure*}
\centering
\includegraphics[width=\textwidth]{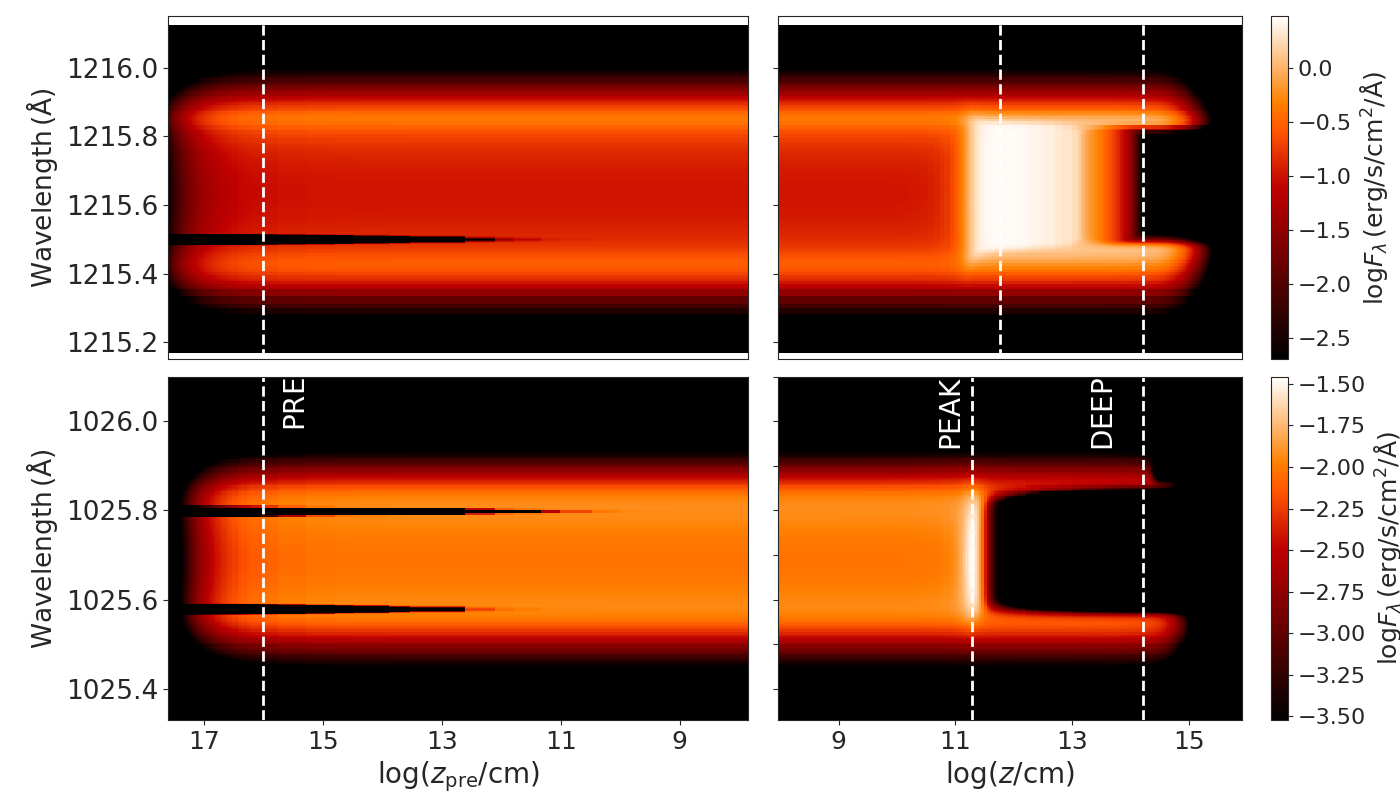}
\caption{Energy density of Ly$\alpha$ (top) and Ly$\beta$ (bottom) emission in the precursor (left) and postshock (right) regions. The vertical dashed lines give the positions of the cuts shown in Fig.~\ref{fig:line_profile}: (PRE) in the precursor at $z_{\rm{pre}}=10^{16}$~cm, (PEAK) at the peaks of Ly$\alpha$ and Ly$\beta$ emission at $z\sim 10^{12}$~cm and $\sim 2\times 10^{11}$~cm respectively, and (DEEP) deeper in the postshock at $z\sim 10^{14}$~cm.} \label{fig:line2d}
\end{figure*}
  
\begin{figure*}
\centering
  \includegraphics[width=0.96\textwidth]{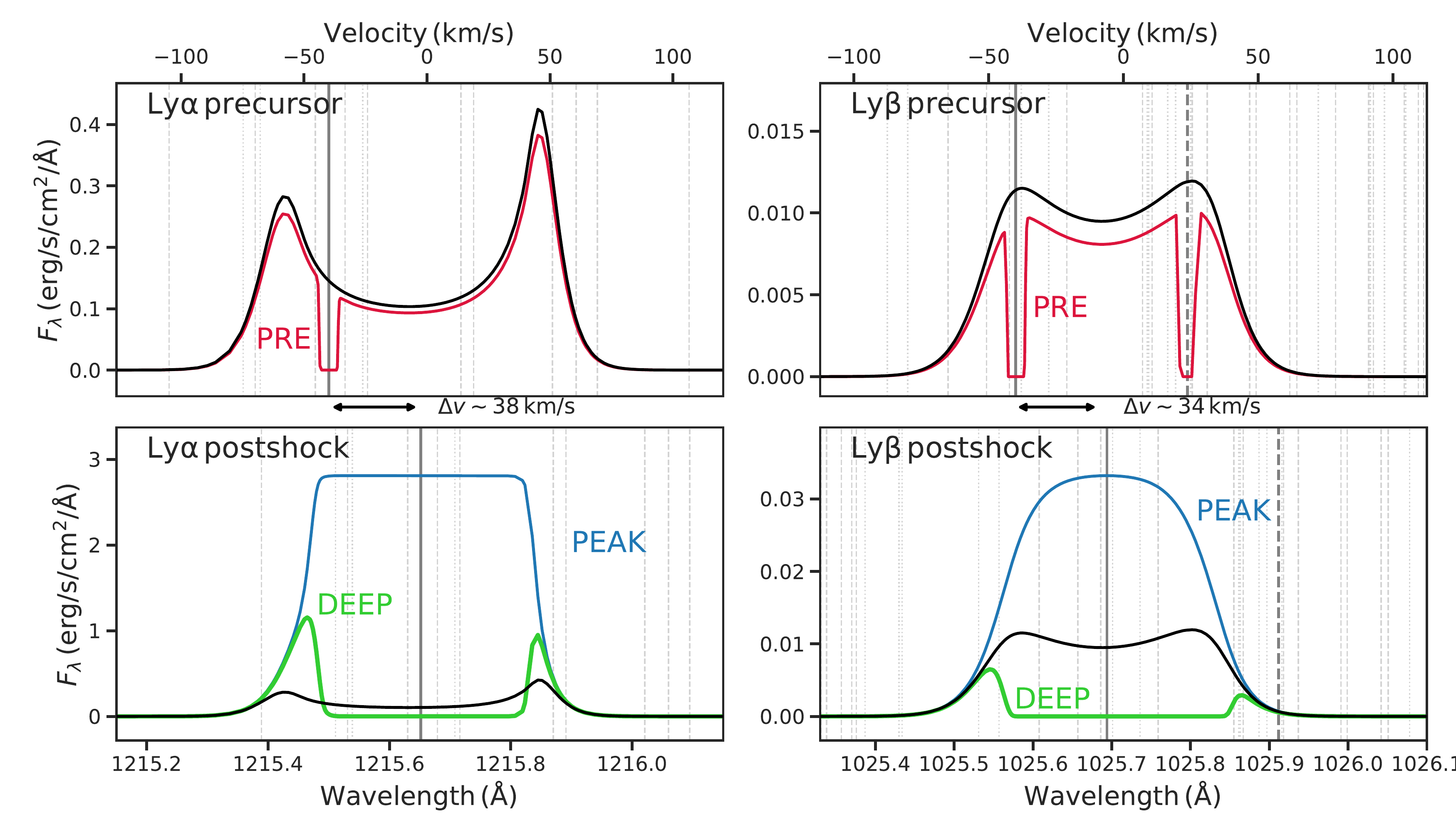}
  \caption{Fluxes of Ly$\alpha$ (left) and Ly$\beta$ (right) in precursor (top) and postshock positions (bottom). The positions of the cuts---labelled PRE (red), PEAK (blue) and DEEP (green)---are shown in Fig.~\ref{fig:line2d} by the vertical dashed lines, and the black lines give the fluxes at the shock front $z=0$. Vertical grey lines show line centres of Ly$\alpha$ and Ly$\beta$ (solid), H$_2$ Lyman (dashed) and Werner (dotted) band absorption lines, with an emphasised $v=6-0$~P(1) H$_2$ line (thick dashed). The percursor vertical lines are Doppler-shifted by the shock velocity, whereas in the postshock panels they are shifted by flow velocity at the peak of the emission, $V_z\sim$4~km/s for Ly$\alpha$ and $V_z\sim$7~km/s for Ly$\beta$. The arrows between the top and bottom panels highlights the velocity shift between the preshock gas and at peak emission. We note that none of the $y$-axes scales are the same.}  
   \label{fig:line_profile}
\end{figure*}

\begin{figure*}
\centering
  \includegraphics[width=\textwidth]{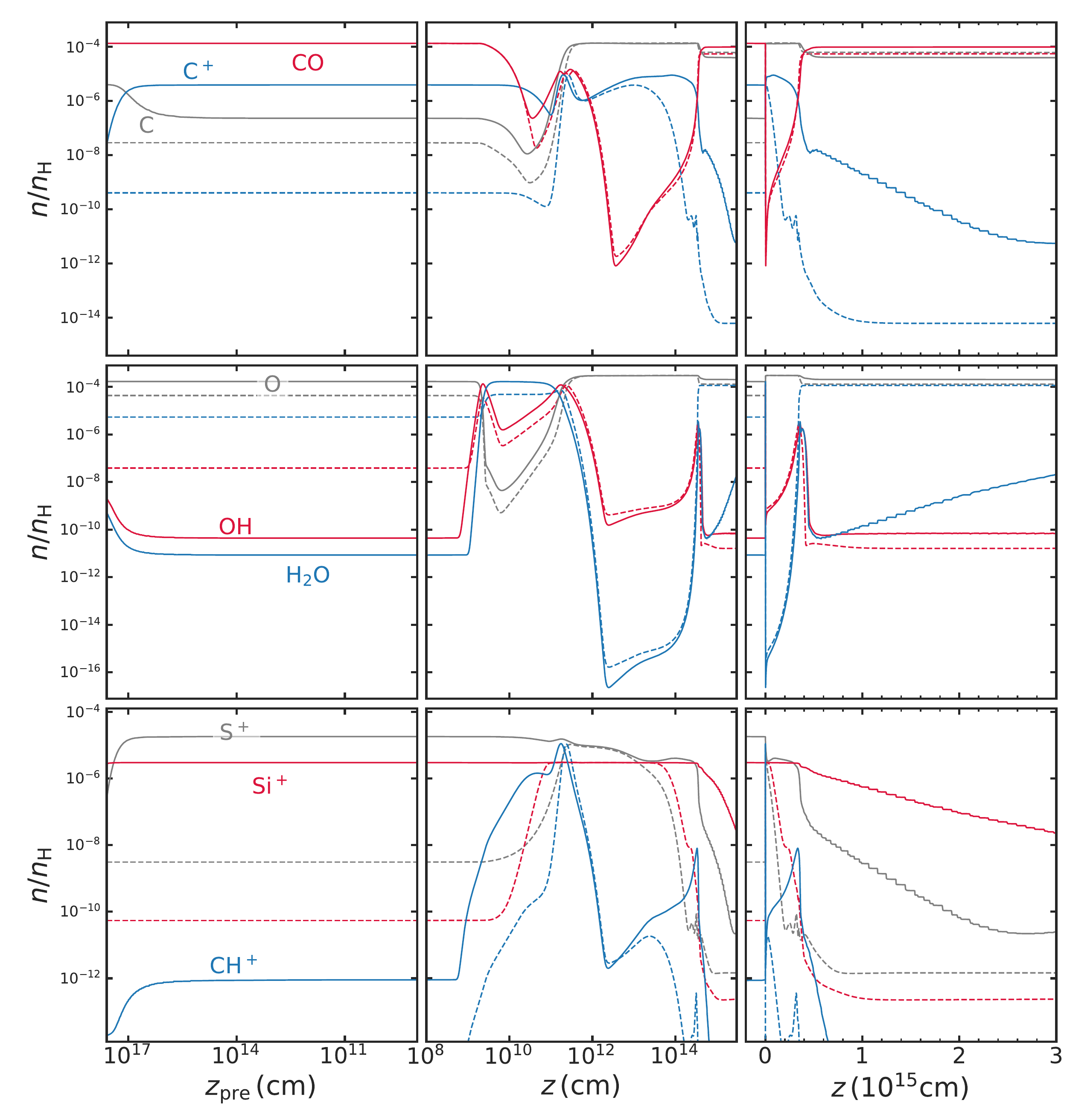}
  \caption{Abundance profiles for selected species for the fiducial shock (Table~\ref{tab:shockparams}) including (solid) and without (dashed) the self-generated UV. The left column shows quantities in the radiative precursor, with distance increasing towards the left (i.e. distance from the shock front), the middle column shows postshock quantities with distance increasing towards the right in log scale while the right column is the same as the middle but in linear scale.}
  \label{fig:profile_C}
\end{figure*}

Profiles of temperature, density, and radiation field for this shock are shown in Fig.~\ref{fig:profile_40} with (solid) and without (thin dashed) the self-generated UV treatment included. The computation typically converges by 3 iterations of shock and radiative transfer cycles. With no radiation field, the shock propagates into cold gas at $\sim$10~K. After the adiabatic jump, seen in the middle panels of Fig.~\ref{fig:profile_40}, dissociation of H$_2$ mostly due to collisions with electrons produces atomic H with abundance $\sim$~1. Cooling due to H excitation by electron collisions---as discussed in section~\ref{sec:Hcooling}---then determines the transition down to a plateau at $T\sim$~10$^4$~K. The cooling is peaked in this transition as the temperature quickly drops too low for significant excitation of H, at which point O becomes the dominant coolant. This plateau is maintained until H$_2$ reforms on dust grains and efficiently cools the gas along with other molecules---such as OH and CO---whose production follows the presence of H$_2$.

With the self-generated UV treatment included, the photons that escape the shock front form a radiative precursor, heating the gas ahead of the shock to $\sim$100~K over a distance of $\sim$10$^{17}$~cm (top left panel of Fig.~\ref{fig:profile_40}), which is the length scale over which dust absorption fully attenuates the field. Photoionisation of C and S produces an increase in electron abundance entering the shock $\sim$~3 orders of magnitude more than the initial abundance. An increase in H$^+$ also generates more electrons, resulting in stronger dissociation of H$_2$ in the adiabatic plateau. Hence the cooling due to H excitation and the transition to the second temperature plateau occurs earlier than without the UV field. The position of this transition converges by the third iteration.

\begin{figure*}
\centering
  \includegraphics[width=\textwidth]{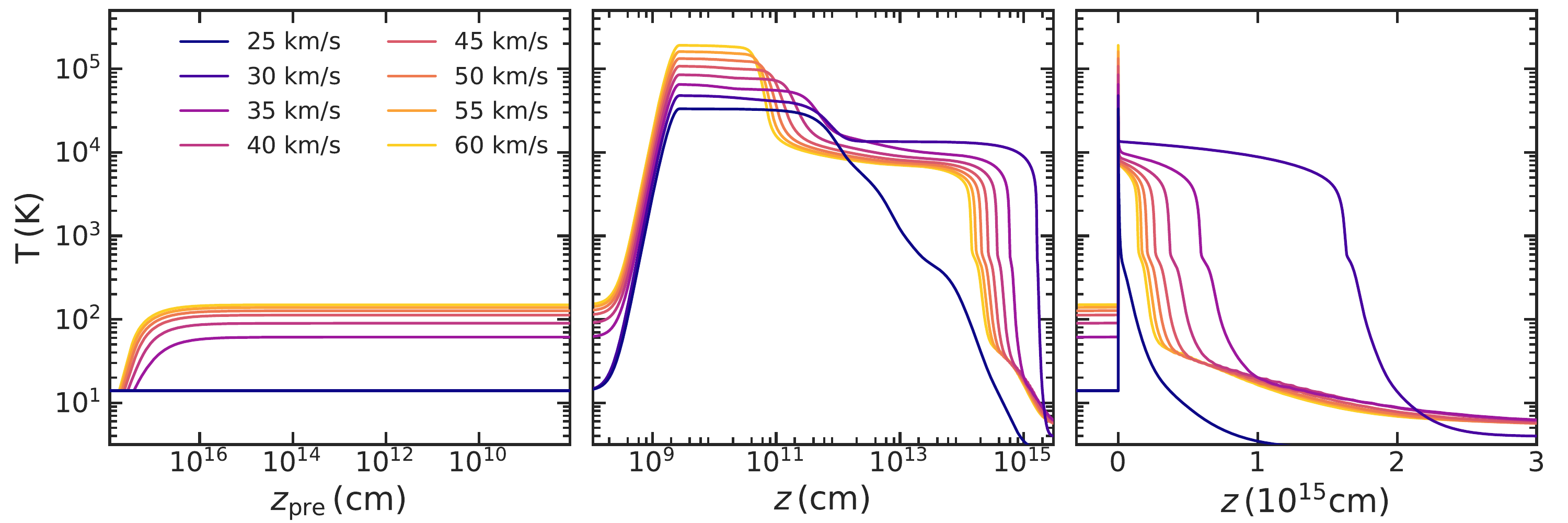}
  \caption{Temperature profiles for the self-consistent shock solutions with shock velocities $V_s=25-60$~km/s propagating into gas at $n_{\mathrm{H}}=10^4$~cm$^{-3}$. The left shows profiles in the radiative precursor, with distance increasing towards the left (i.e. distance from the shock front), the middle shows postshock profiles with distance increasing towards the right in log scale while the right is the same as the middle but in linear scale. Peak temperatures increase with increasing shock velocity (see Eq.~\ref{eq:rankineT}).}
  \label{fig:n4_temps}
\end{figure*}

Strong self-absorption traps most of the Ly$\alpha$ photons near a peak of emission, seen in the lower centre panel of Fig.~\ref{fig:profile_40}, between $z=10^{11}$--$10^{14}$~cm where the UV flux is an order of magnitude more intense ($G_{\mathrm{eff}}\sim$~450) than that which escapes ($G_{\mathrm{eff}}\sim$~55).  Emerging from the shock front, the Ly$\alpha$, Ly$\beta$, and two-photon fluxes are roughly 39, 1, and 14 times the ISRF. The flux escaping the trapped region is not symmetric. For Ly$\alpha$, the rightward escaping photon flux is $\sim$14\% more than leftward. The rightward escaping photons are attenuated by dust in the postshock, generating an extended tail of warm molecular gas due to photoelectric heating. This tail remains above 10~K for a factor of $\sim$3 longer than the first iteration.
    
Figures~\ref{fig:line2d} and \ref{fig:line_profile} show the radiation field in more detail. Figure~\ref{fig:line2d} gives the flux around Ly$\alpha$ and Ly$\beta$ wavelengths at all positions in the precursor and postshock regions. It clearly shows strong absorption in the line cores forcing the energy to escape in the line wings.  Figure~\ref{fig:line_profile} shows a few representative spectra at the shock front ($z=0$), in the precursor ($z_{\rm{pre}}=10^{16}$~cm), at the peak of Ly$\alpha$ ($z\sim 10^{12}$~cm) and Ly$\beta$ ($z\sim 2\times 10^{11}$~cm) emission, and deep into the postshock ($z=10^{14}$~cm). The peaks of the wings in the precursor are shifted further away from line centre than the wings in the deep postshock after the peak of emission. This is because the H in the fluid before the peak is at higher temperatures than after the peak, and so it has a wider absorption profile. The precursor spectrum at Ly$\alpha$ shows deep absorption due to the cold H in this region at negative of the shock velocity. This is also seen in the Ly$\beta$ profile, as well as a deep absorption line due to H$_2$~$v=6-0$~P(1) Lyman band absorption, whose line centre Doppler-shifted by fluid velocity is shown in the Ly$\beta$ panels of Fig.~\ref{fig:line_profile} by the dashed grey vertical lines. The spectrum at the peak of Ly$\alpha$ emission shows the typical flat-top profile due to saturation effects of optically thick emission. The same saturation effects make the Ly$\beta$ spectrum start to show the flat-top profile. These peaks of emission emerge from the hot regions of the shock at T$\sim 50000$~K.

Profiles of selected carbon-, oxygen-, sulphur-, and silicon- bearing species are shown in Fig.~\ref{fig:profile_C}. The shock-generated UV strongly changes the preshock abundances and ionisation fraction of C, S, and Si. Without the UV field these 3 species all enter the shock mostly neutral, whereas they all become mostly ionised in the preshock. Oxygen chemistry is also strongly affected, with the photodissociation cross-sections of oxygen-bearing molecules O$_2$, H$_2$O, and OH overlapping Ly$\alpha$, Ly$\beta$, and/or the two-photon continuum. On the other hand, the photodissociation of CO proceeds via the indirect predissociation mechanism, requiring line absorption at specific wavelengths that do not overlap the H emission \citep{van_dishoeck_photodissociation_1988}. Hence as the gas begins to be shocked most of the oxygen is contained in atomic O and CO. H$_2$ is another molecule that survives the strong UV radiation, however in this case it is because there is not enough time spent in the radiative precursor for strong photodissociation to take place. When the fluid enters the hot shock front the temperatures and densities of the adiabatic jump and transition to the second plateau are not so different with or without the UV field, so the chemistry produces very similar abundances by $z\sim$10$^{12}$~cm. However, the UV persists with fluxes stronger than the typical ISRF, that is $G_{\rm eff}>1$, until $z\sim$10$^{15}$~cm strongly affecting the chemistry in the region $z\gtrsim$10$^{13}$~cm. For example, the atomic ions C$^+$, S$^+$, and Si$^+$ all remain more than 2 orders of magnitude higher than when UV photo-ionisation is not taken into account. Oxygen remains mostly atomic whereas without the UV treatment there is a very strong formation of H$_2$O in the this region. Finally, the peak of CH$^+$ production at $z\sim$10$^{14}$~cm gives an abundance 3-4 orders of magnitude larger than without the UV treatment, before it settles at close to the same level after the UV has been attenuated. We address in the following sections how these differences in local abundances affect integrated quantities such as line emission and column densities.


\subsection{Intermediate velocity shocks: $V_s=25$--$60$~km/s}\label{sec:grid}

\subsubsection{Critical velocity}

As a first step in determining the velocity at which the self-generated UV becomes important to treat we computed a small grid of shock models with velocities $V_s=25$--$60$~km/s at every 5~km/s, propagating into gas with total hydrogen density $n_{\rm{H}}=10^4$~cm$^{-3}$. All other parameters are the same as the fiducial shock, listed in Table~\ref{tab:shockparams}. We leave the exploration of the dependence on density and other parameters to a forthcoming work.

\begin{figure}
\centering
  \includegraphics[width=\columnwidth]{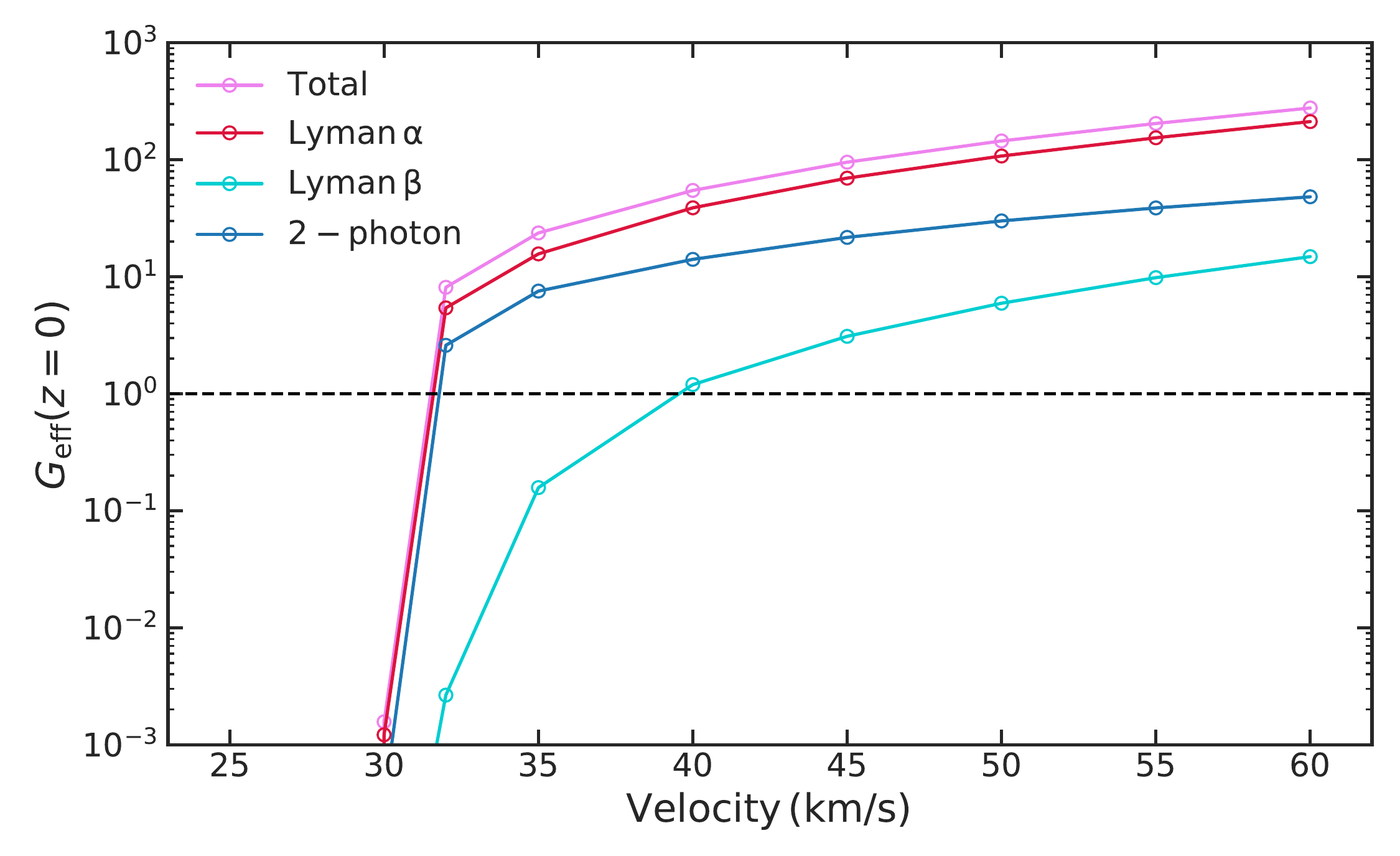}
  \caption{Total UV flux as computed with equation~\ref{eq:geff} (violet) and the contributions to this total by Ly$\alpha$ (red), Ly$\beta$ (cyan), and two-photon continuum (blue) at the shock front ($z=0$) for shocks with velocities $V_S=25-60$~km/s and preshock densities $n_{\rm{H}}=10^4$~cm$^{-3}$. The dashed horizontal line represents the standard interstellar radiation field.}
  \label{fig:photons}
\end{figure}

The self-consistent converged temperature profiles are shown in Fig.~\ref{fig:n4_temps}. The peak temperature of the adiabatic jump rises with velocity according to Eq.~\eqref{eq:rankineT}. At $V_s=25$~km/s the peak temperature is not strong enough to dissociate H$_2$ to remove it as a significant coolant, and so the temperature drops to the postshock equilibrium without any plateaus. As the shock velocity increases, the dissociation occurs faster due higher temperatures. Thus atomic H is produced earlier with increasing velocity and so the transition to the second plateau occurs earlier due to the cooling by collisional excitations of H with electrons. Stronger H cooling leads to larger fluid densities in the second plateau. In turn, this leads to larger dust densities and a faster H$_2$ formation rate, so that H$_2$ becomes the dominant coolant and the second plateau ends earlier with increasing velocity. The attenuation due to enhanced dust densities balances the stronger UV radiation fields here to give a nearly constant distance at which the shocks cool down to, say, 10~K.

The temperature profiles of the radiative precursors---left panel of Fig.~\ref{fig:n4_temps}---show that heating due to the self-generated UV field starts to increase the preshock temperature above the initial value at a velocity between 30 and 35 km/s. We ran a shock at 32~km/s to sample this transition with more detail. The UV strength parameter $G_{\mathrm{eff}}$ emerging from the shock front (z=0) is shown in Fig.~\ref{fig:photons}. This shows a dramatic transition in the shock velocity range $V_s = 30$--32~km/s at which the emergent UV photon flux is stronger than the ISRF, that is $G_{\mathrm{eff}}>1$. The adiabatic jump at this velocity heats the gas to high enough temperatures to simultaneously remove H$_2$ as a strong coolant via collisional dissociation and excite H enough to produce strong Ly$\alpha$ fluxes. At $V_s=35$~km/s the Ly$\alpha$ photon flux alone is $\sim$~10 times the ISRF. This presents the rather unique situation of having a slab of fully molecular material up against a boundary radiation field with $G_{\mathrm{eff}}$=10--400 for shocks with velocities between 35--60~km/s. This situation is generally not realised in photon-dominated regions with equivalent UV flux, because the broad spectrum of the ISRF is more effective at photodissociating H$_2$ than the shock UV field concentrated around Ly$\alpha$ and Ly$\beta$ wavelengths. In such slabs molecule formation thus takes place once the UV field has been strongly attenuated.

\subsubsection{Shock length- and time-scales} \label{sec:sizes}
  
\begin{figure}
\centering
\includegraphics[width=\columnwidth]{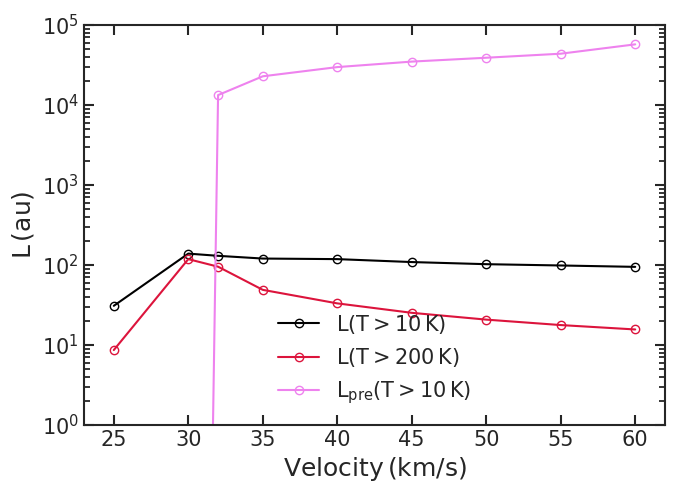}
\caption{Lengths at which the shocked regions cool down 10~K (black), or down to 200~K (red), and size of the radiative precursor computed as the gas layer ahead of the shock heated above 10~K (violet).}
\label{fig:shocksizes}
\end{figure}

\begin{figure}
\centering
\includegraphics[width=\columnwidth]{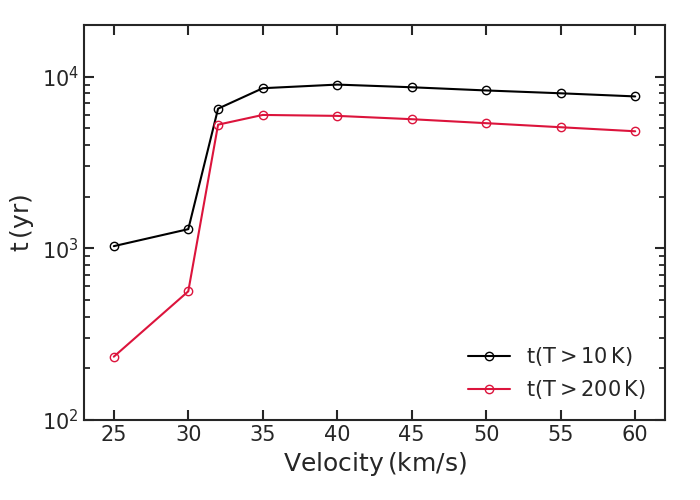}
\caption{Shock timescales for the same criteria as Fig.~\ref{fig:shocksizes}.}
\label{fig:shocktimes}
\end{figure}

In Figs.~\ref{fig:shocksizes} and \ref{fig:shocktimes} we show the shock sizes and timescales at which the shock has cooled down to 200~K or 10~K. For the size scale, we also plot the size of radiative precursor that has been heated above 10~K. The sizes of both radiative precursor and postshock region are somewhat constant for shock velocities $V_s\geq 35$, varying by less than a factor of 2 depending on the size criteria. This is because these sizes are essentially determined by the dust attenuation of the UV field. 

\subsubsection{Line emission}\label{sec:lines}

\begin{figure}
\centering
\includegraphics[width=\columnwidth]{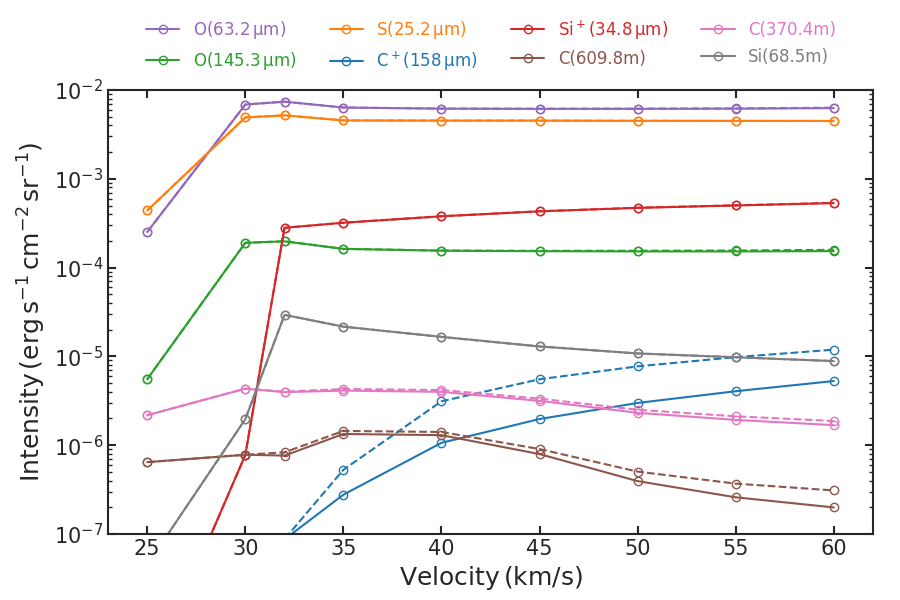}
\caption{Selected atomic fine-structure line intensities generated in the post-shock region (solid) or in the radiative precursor in addition to the post-shock (dashed).}
\label{fig:finestructure}
\end{figure}

\begin{figure}
\centering
\includegraphics[width=\columnwidth]{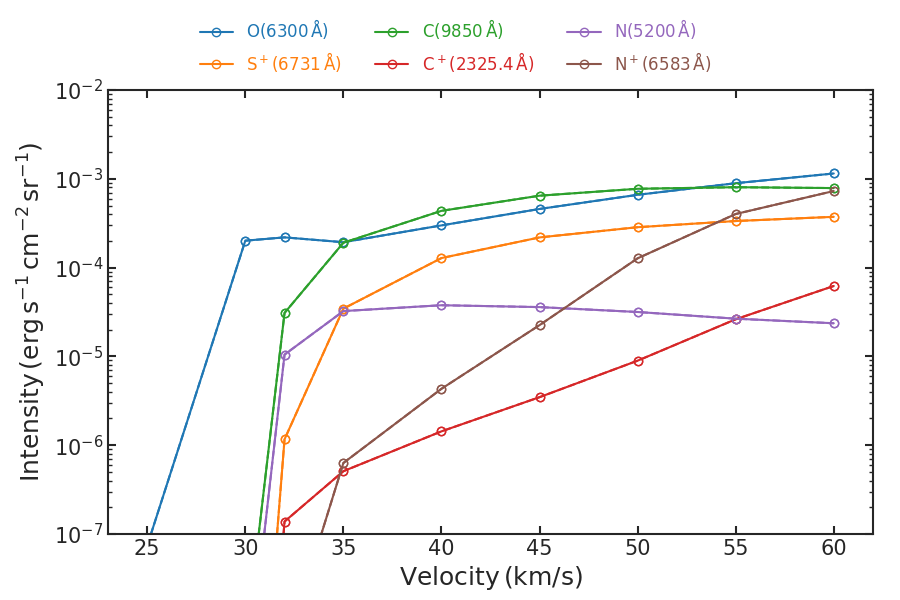}
\caption{Selected atomic meta-stable line intensities generated in the post-shock region (solid) or in the radiative precursor in addition to the post-shock (dashed).}
\label{fig:metastable}
\end{figure}

\begin{figure}
\centering
\includegraphics[width=\columnwidth]{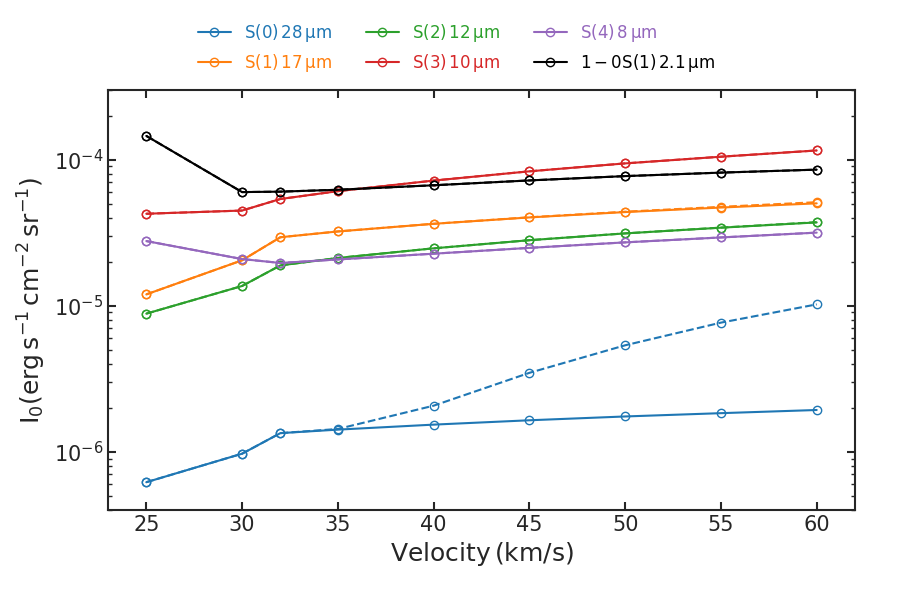}
\caption{Selected ro-vibrational H$_2$ line intensities generated in the post-shock region (solid) or in the radiative precursor in addition to the post-shock (dashed).}
\label{fig:h2lines}
\end{figure}

The shock observables that we consider here, that is selected atomic emission lines, rovibrational lines of H$_2$, and column densities of various species, are all quantities obtained by integrating local quantities through the shock. The choice of where to end that integration therefore determines these quantities and the interpration of observations. For this work we integrate until the gas falls down to 10~K, so that we integrate through shock-heated gas. There is also material ahead of the shock heated above 10~K in the radiative precursor. However, the precursor slabs computed here assume a uniform medium over large distances, which is not necessarily relevant in real astrophysical systems. Because it is unknown how realistic it is to include emission generated in the precursor, so we simply consider two cases: including this material or not.

In Fig.~\ref{fig:finestructure} we show the intensity of selected infrared fine structure lines generated in the shock  heated gas. Most of these lines show a dramatic increase in intensity over the velocity range 30--60~km/s compared to lower velocity shocks. For example, the C$^+$(158$\mathrm{\mu}$m) and Si$^+$(34.8$\mathrm{\mu}$m) lines shows increases of more than 2 orders of magnitude once the UV becomes important. The C$^+$ line also shows an increasing trend with velocity in this range, and has a strong contribution from the gas heated in the radiative precursor. The other fine structure lines show a remarkably constant intensity in this range. This figure then shows the general result that intermediate velocity molecular shocks produce strong emission in the fine-structure lines O(63.2~$\mu$m), O(145.3~$\mu$m), S(25.2~$\mu$m), and Si$^+$(34.8~$\mu$m).

Intensities of selected metastable atomic lines are shown in Fig.~\ref{fig:metastable}. More optical lines are accounted for in the model, but we show just the strongest lines for each species. As with the fine structure lines there are dramatic increases in these intensities over lower velocity shocks. In addition, there are stronger trends with velocity, for example the N$^+$(6583~$\AA$) line varies by more than 3 orders of magnitude over the velocity range. A general result is that the intermediate velocity molecular shocks produce strong emission in the metastable lines O(6300~$\AA$), C(9850~$\AA$), and S$^+$(6731~$\AA$).

In Fig.~\ref{fig:h2lines} we plot the intensities of the pure rotational lines S(0) up to S(4) of H$_2$ as well as the $v=1-0$~S(1) generated in the shock-heated gas. Unlike the atomic lines, these lines do not show a significant increase compared their intensity from the 25~km/s shock. They are also remarkably constant over the velocity range, except for the contribution to the S(0) by the radiative precursor. Hence combined observations of atomic lines and H$_2$ lines would be necessary to probe the different shock velocities in systems with an ensemble of shocks, such as in the turbulent cascade in the wakes galactic outflows or supernova shocks. These H$_2$ lines lie in the observational bands of JWST and so could be used to interpret planned observations of such astrophysical systems.

\begin{table}
\caption{Energy flux (erg/s/cm$^2$) emerging from the shock front in atomic H lines. The shock kinetic flux is shown for comparison.}
\begin{tabular}{c|ccc|c} \cline{1-5}
\multicolumn{5}{c}{\vspace{-0.3cm}} \\ 
$V_s\,\rm{(km/s)}$ & $\rm{Ly}\alpha$ & $\rm{Ly}\beta$ & $\rm{2ph}$ & $\rm{Kinetic}$ \\ \cline{1-5}
\multicolumn{5}{c}{\vspace{-0.3cm}} \\ 
25 & 0        & 0        & 0           & 0.18 \\
30 & 2.43(-06) & 5.90(-07) & 1.19(-08) & 0.32 \\
35 & 3.97(-02) & 1.30(-02) & 4.73(-04) & 0.50 \\
40 & 9.86(-02) & 2.46(-02) & 3.59(-03) & 0.75 \\
45 & 1.76(-01) & 3.79(-02) & 9.30(-03) & 1.07 \\
50 & 2.73(-01) & 5.25(-02) & 1.78(-02) & 1.47 \\
55 & 3.90(-01) & 6.77(-02) & 2.95(-02) & 1.96 \\
60 & 5.37(-01) & 8.43(-02) & 4.46(-02) & 2.54 \\ \cline{1-5}
\end{tabular}\label{tab:emissionH}

\emph{Note}. Numbers in parentheses denote powers of ten.
\end{table}

The UV emission from atomic H---Ly$\alpha$, Ly$\beta$ or two-photon continuum---are also possible observables. In Table~\ref{tab:emissionH} we list their fluxes that escape into the radiative precursor as well as the shock kinetic flux  $0.5 \rho V_s^3$. For shocks with velocities $V_s \geq 40$~km/s, 13-21\% of the kinetic energy entering the shock comes back ahead of the shock in Ly$\alpha$, Ly$\beta$ or two-photon emission. This UV emission can be absorbed by dust in the material ahead of the shock, which is often an unknown quantity in astrophysical systems. Combined with the previously discussed shock tracers,  Table~\ref{tab:emissionH} could then be used to give a prediction of the maximum contribution to the Ly$\alpha$ emission from intermediate velocity shocks.

\subsubsection{Column densities}

\begin{figure*}
\centering
  \includegraphics[width=\textwidth]{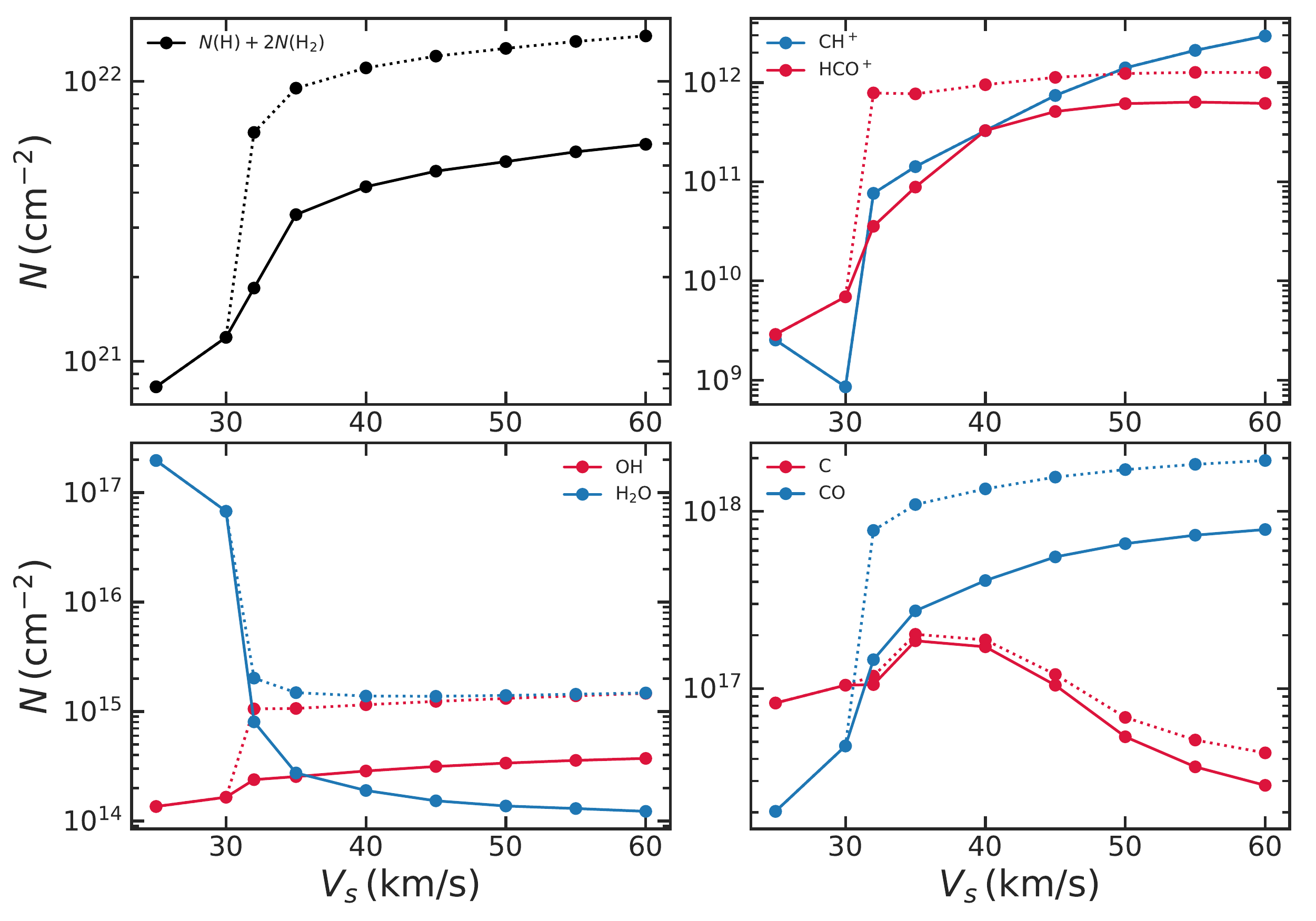}
  \caption{Column densities of selected species. In each panel we show the column density for postshock regions above 10~K (solid) and with the heated material in radiative precursor added to the postshock (dotted).}
  \label{fig:column_dens1}
\end{figure*}

In addition to atomic and H$_2$ line emission there are shock tracers in the molecular chemistry. Column densities of selected species are shown in Fig.~\ref{fig:column_dens1}. We show the column densities for postshock regions integrated until the gas cools to 10~K (solid lines) and also with the material in the radiative precursor added to shocked gas (dotted). Due to the extended warm tail produced by the UV there is an increase in the total H column density---by a factor $\sim$3--4 in the postshock or an order of magnitude when including the preshock---for shocks with velocities $V_s>30$~km/s. Above this velocity, however, there is no significant correlation with velocity. This is due to the roughly constant shock-size, discussed in the previous section, over the velocity range once Ly$\alpha$ production becomes significant.  

Some species show enhanced column densities when the UV field becomes strong at velocities $V_s \geq 32$~km/s. For example, column densities of CH$^+$ and HCO$^+$ are increased by more than 2 orders of magnitude by the end of the velocity range compared to their values at $V_s=30$~km/s. This is due to the larger abundances of C$^+$ maintained by photoionisation deep in the postshock. The updated UV treatment may therefore allow us to consider shock models to interpret extragalactic ALMA observations of CH$^+$ emission \citep{falgarone_2017} as well as protostellar jet observations of HCO$^+$ \citep[e.g.][]{tafalla_molecular_2010}. In the other direction, photodissociation of H$_2$O strongly reduces its column densities compared to molecular shocks at lower velocities. Observations involving oxygen chemistry have long stimulated the development of shock models. For instance, recent \textit{Herschel} observations of protostellar environments reveal H$_2$O abundances far too low to be explained with simple chemical models (e.g. \citealt{goicoechea_complete_2012,kristensen_observational_2013,karska_shockingly_2014,karska_herschel_2018}). The precise origins of the emission of O$_2$ (e.g. \citealt{yildiz_deep_2013,chen_herschel_2014,melnick_o2_2015}) and O \citep{kristensen_oxygen_2017} have met with similar struggles. Mixtures of high velocity dissociative shocks, photodissociation regions, low velocity C-type shocks, and shocks irradiated by UV from either external sources or the shocks themselves have been invoked to explain these observations. The present work broadens the range of parameters of the Paris-Durham shock code to be used to study these problems, and shows how self-irradiated shocks could be a viable solution to explain low abundances of H$_2$O and O$_2$. The column densities of all these species (except CH$^+$) in the heated precursor are larger than in the postshock because the precursors are orders of magnitude larger than the postshock, as seen in Fig.~\ref{fig:shocksizes}. H$_2$O, OH, CO, and total H have more than a factor of 3 larger column in the radiative precursor than in the shocked gas. In summary, molecular shocks at intermediate velocities ($V_s=$32--60~km/s) have large column densities of CH$^+$, HCO$^+$, and CO while also having low column densities of H$_2$O.

In Appendix~\ref{app:neufeld_comparison} we compare the column densities of selected species in $V_s=60$~km/s shocks to the results of the shock models of \cite{neufeld_fast_1989}. The good agreement between the two works is a striking result given the decades of updates to chemical reaction networks and rates, computational methods, and the inclusion of the magnetic field in our work.
  
\section{Conclusion}\label{sec:conclusion}
We have implemented a treatment of the UV radiative transfer including the atomic hydrogen lines Ly$\alpha$, Ly$\beta$, and the two-photon continuum in order to solve for self-irradiated molecular shocks at intermediate velocities using the Paris-Durham public shock code. The main results are summarised as follows:
\begin{itemize}
\item A detailed treatment of radiative transfer is necessary to accurately compute the line profiles and escape of Ly$\alpha$ and Ly$\beta$. However, to understand the energetic impacts of H excitation in these shocks it is sufficient to model the radiation with an optically thin treatment.
\item For preshock density $n_{\rm{H}}=10^4$~cm$^{-3}$, shocks with velocity $V_s > 30$~km/s produce a UV radiation field that escapes into the preshock gas with a Ly$\alpha$ photon flux stronger than the standard ISRF.
\item For shock velocities between 35--60~km/s the escaping UV photons give a radiation field parameter $G_{\rm eff}\sim$~10--400. In other words, the line and continuum emission of H alone carries as much as 400 times more photons than those contained in the entire ISRF. These photons escaping ahead of the shock heat the preshock gas in a radiative precursor over $\sim$~10$^{17}$~cm to $\sim$~100~K.
\item The maximum contribution to Ly$\alpha$ emission for shocks with velocities 40--60~km/s is as large as 13-21\% of the shock kinetic energy flux.
\item After HI emission, molecular emission is the second most important coolant in intermediate velocity shocks. In addition, these shocks have large column densities of CH$^+$, HCO$^+$, and CO while also having low column densities of H$_2$O. These shocks may be useful in explaining low H$_2$O abundances found in \textit{Herschel} observations of protostellar environments.
\item Compared to lower velocity shocks, atomic fine structure and metastable emission lines are boosted by many orders of magnitude. For example, the O(63.2~$\mu$m) and S(25.2~$\mu$m) fine-structure lines and O(6300~$\AA$) and C(9850~$\AA$) metastable lines are particularly bright. Most fine structure lines are remarkably constant at intermediate shock velocities, but ratios of metastable lines could be used as a probe of shock velocity.
\item We predict intensities of ro-vibrational lines of H$_2$. Many of these lines will be observable by \textit{JWST} and hence this treatment will be critical for the interpretation of observations of phenomena with shocks at these velocities.
\end{itemize}
The present work shows that it is indeed important to properly model the UV radiation generated by the shock heated gas in order to extend models of molecular shocks to higher velocities. A forthcoming work will present the application of the present models for the interpretation of observations.

\section*{Acknowledgments}
This research has received funding from the European Research Council through the Advanced Grant MIST (FP7/2017-2022, No 742719). We would also like to acknowledge the support from the Programme National “Physique et Chimie du Milieu Interstellaire” (PCMI) of CNRS/INSU with INC/INP co-funded by CEA and CNES.

\bibliographystyle{aa}
\bibliography{selfirrad}

\clearpage
\appendix

\section{Number of H$_2$ levels}\label{app:h2levels}

\begin{figure}[h]
  \centering
  \includegraphics[width=\columnwidth]{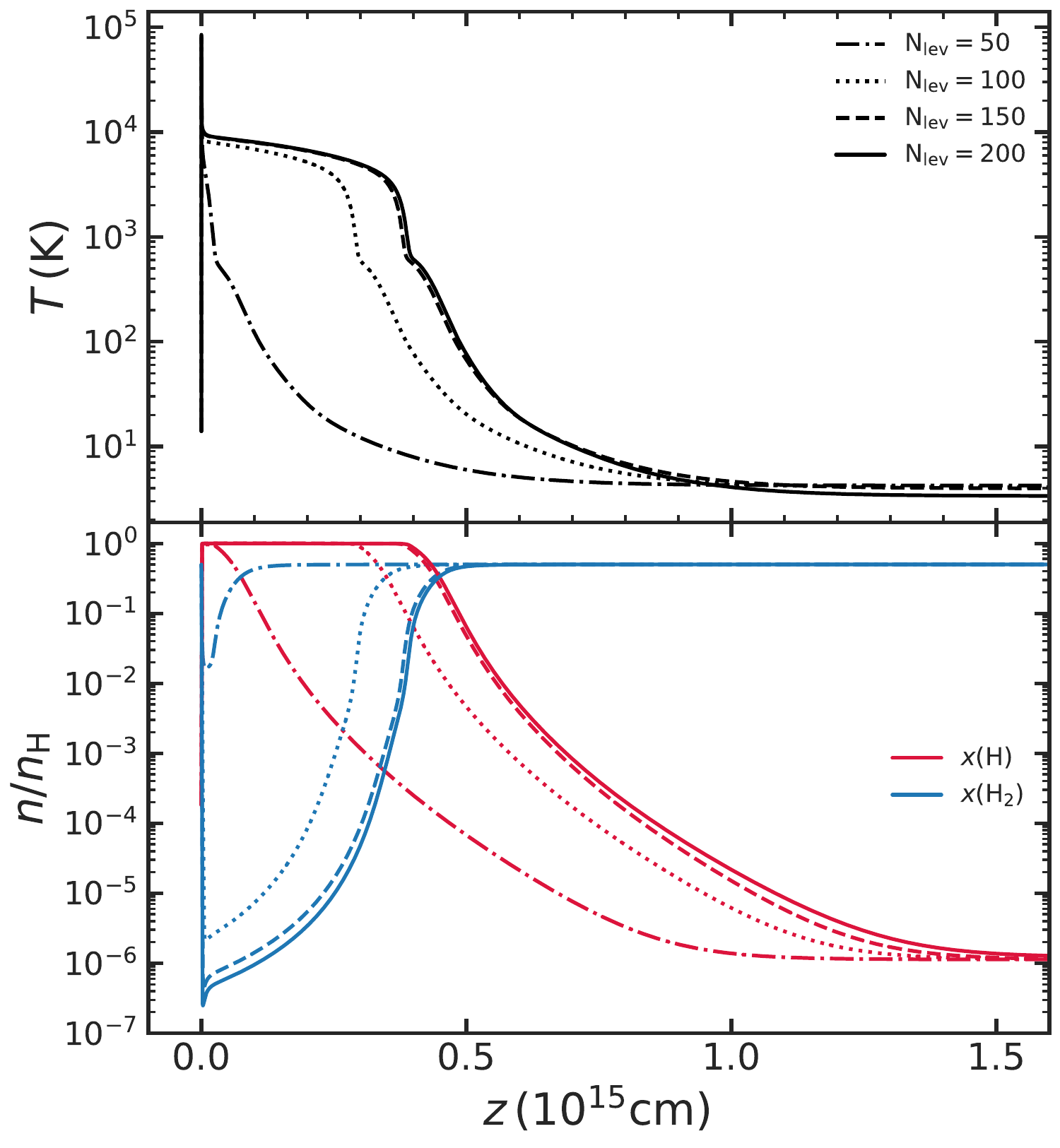}
  \caption{Profiles of temperature (top) and abundances of H and H$_2$ (bottom) for $V_s=40$~km/s shocks computed treating 50 (dot-dashed), 100 (dotted), 150 (dashed), or 200 (solid) levels of H$_2$. }
  \label{fig:h2levels}
\end{figure}

To estimate the impact of the number of H$_2$ levels on the shock structure we run 4 shock models with shock velocity $V_s=40$~km/s and preshock density $n_{\rm H}=10^4$~cm$^{-3}$, varying the number of H$_2$ levels treated: N$_{\rm lev}$=50, 100, 150 or 200. Otherwise, we use the same input parameters as the fiducial model (see Table~\ref{tab:shockparams}). In Fig.~\ref{fig:h2levels} we show the temperature and hydrogen abundance profiles for these shocks, which are reasonably converged when 150 levels are included. When treating 50 levels the dissociation of H$_2$ is underpredicted by $\sim 4$ orders of magnitude, leading to the reformation of H$_2$ to occur earlier and reduce the size of the shock by a factor of $\sim 5$.

\begin{figure}
  \centering
  \includegraphics[width=\columnwidth]{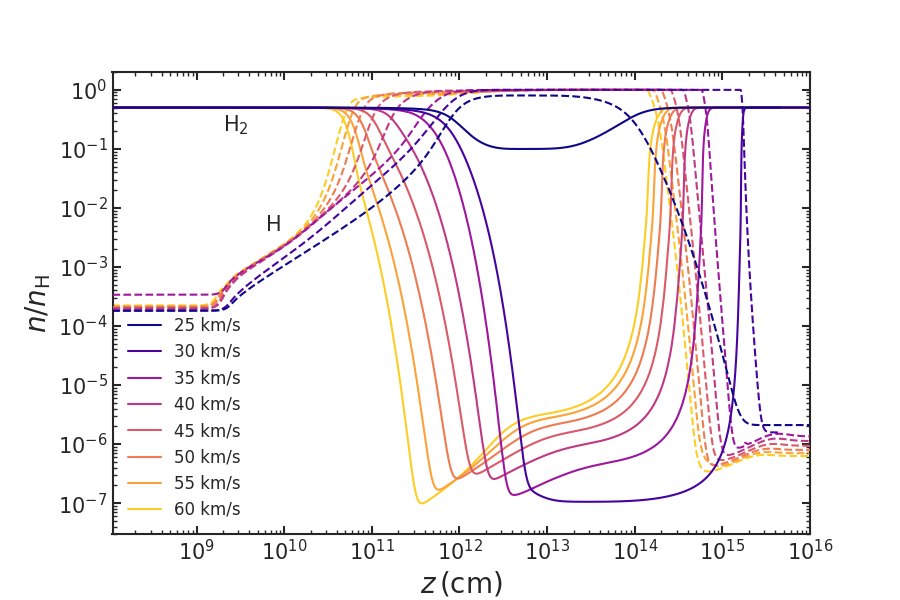}
  \caption{Profiles of H and H$_2$ abundance for shocks with velocities $V_s=25$--60~km/s and preshock density $n_{\rm H}=10^4$cm$^{-3}$.}
  \label{fig:hydrogen_profiles}
\end{figure}

Using 150 levels of H$_2$, Fig.~\ref{fig:hydrogen_profiles} shows abundances profiles of H$_2$ and H over the grid of shock models described in section~\ref{sec:grid}. At $V_s$=25~km/s, H$_2$ is strongly dissociatied in the adiabatic jump and H is the dominant species, but there is a transition at $V_s$=30~km/s above which shocks dissociate H$_2$ to a similar abundance $x({\rm H}_2)\sim$10$^{-7}$.

\section{Atomic hydrogen parameters}\label{app:hydrogen}

The three level model of hydrogen described in section~\ref{sec:atomH} is built by combining the data of the sub-levels with energies and statistical weights taken from the NIST database \citep{NIST_ASD} and listed in Table~\ref{tab:Hlevels}. The statistical weights of the combined levels are the sum of the weights of their sub-levels   \begin{align}
g_n = \sum_{i\in n} \hat{g}_i. 
\end{align}
The energies of the combined levels are the average energies of their sub-levels weighted by the statistical weights
\begin{align}
E_n = \frac{1}{g_n}\sum_{i\in n} \hat{g}_i \hat{E}_i.
\end{align}
Similarly the Einstein coefficients of the combined levels, listed in Table~\ref{tab:Hparams}, are the weighted average Einstein coefficients
\begin{align}
A_{ul} = \frac{1}{g_u} \sum_{i,j} \hat{g}_i \hat{A}_{ij}
\end{align}
where the sum is over all the transitions from the sub-levels of the upper combined level to any of the sub-levels of the lower combined level. The sub-level transition data are taken from the NIST database \citep{NIST_ASD}---with the exception of the two-photon Einstein coefficient taken from \cite{nussbaumer_1984}---and listed in Table~\ref{tab:Hradtrans}.
  
\begin{table}
\centering
\caption{Sub-level data of atomic hydrogen combined into 3-level model taken from the NIST database  \citep{NIST_ASD}.}
\begin{tabular}{c c c c c c} \cline{1-6}\cline{1-6}
 \multicolumn{6}{c}{\vspace{-0.3cm}} \\ 
 i & n & $\hat{g}_i$ & $\hat{E}_i$ (K) & Orbital ($nl$) & $j$\\ \cline{1-6}
 \multicolumn{6}{c}{\vspace{-0.3cm}} \\ 
 1 & 1 & 2 & 0 & 1s & 1/2\\
 2 & 2 & 2 & 118352.24454327 & 2p & 1/2\\
 3 & 2 & 2 & 118352.29531193 & 2s & 1/2\\
 4 & 2 & 4 & 118352.77097467 & 2p & 3/2\\
 5 & 3 & 2 & 140269.55551305 & 3p & 1/2\\
 6 & 3 & 2 & 140269.57062165 & 3s & 1/2\\
 7 & 3 & 4 & 140269.71123621 & 3d & 3/2\\
 8 & 3 & 4 & 140269.71149231 & 3p & 3/2\\
 9 & 3 & 6 & 140269.76322354 & 3d & 5/2\\ \cline{1-6}
\end{tabular}\label{tab:Hlevels}
\end{table}
  
\begin{table}
\centering
\caption{Radiative transition data for transitions between sub-levels of atomic hydrogen, taken from the NIST database \citep{NIST_ASD} except for two-photon Einstein coefficient (transition 9).}
  \begin{tabular}{c c c c l r} \cline{1-5}\cline{1-5}
  \multicolumn{5}{c}{\vspace{-0.3cm}} \\ 
    idx & i & j & $\hat{E}_{ij}$ (K) & $\hat{A}_{ij}$ (s$^{-1}$) \\ \cline{1-5}
  \multicolumn{5}{c}{\vspace{-0.3cm}} \\
	1 & 6 & 4  &  21916.79964698 & 4.2097(+06) \\                           
	2 & 7 & 4  &  21916.94026154 & 1.0775(+07) \\                           
	3 & 9 & 4  &  21916.99224887 & 6.4651(+07) \\                           
	4 & 5 & 3  &  21917.26020112 & 2.2449(+07) \\                           
	5 & 6 & 2  &  21917.32607838 & 2.1046(+06) \\                           
	6 & 8 & 3  &  21917.41618038 & 2.2448(+07) \\                           
	7 & 7 & 2  &  21917.46669294 & 5.3877(+07) \\                           
	8 & 2 & 1  & 118352.24454327 & 6.2649(+08) \\                           
	9 & 3 & 1  & 118352.29531193 & 8.2000(+00) \\                           
	10 & 4 & 1 & 118352.77097467 & 6.2648(+08) \\                           
	11 & 5 & 1 & 140269.55551305 & 1.6725(+08) \\                           
	12 & 6 & 1 & 140269.57062165 & 1.1090($-06$) \\                           
	13 & 7 & 1 & 140269.71123621 & 5.9380(+02) \\                           
	14 & 8 & 1 & 140269.71149231 & 1.6725(+08) \\                           
	15 & 9 & 1 & 140269.76322354 & 5.9370(+02) \\ \cline{1-5}
\end{tabular}\label{tab:Hradtrans}

\emph{Note}. Numbers in parentheses denote powers of ten.
\end{table}

The effective collision strengths for de-excitation due to collisions with electrons are taken from \cite{anderson_2002}, listed in Table~\ref{tab:Hcolstrength}. The effective collision strengths, $\Upsilon_A$, have been modified in order to be applied to the fine structure of H. For all transitions $n'l'j' \to nlj$
\begin{align}
\hat{\Upsilon} = \hat{\Upsilon}_A\left( n'l \to nl \right) \frac{(2j'+1)}{(2s'+1)(2l'+1)} \frac{(2j+1)}{(2s+1)(2l+1)}.
\end{align}
The de-excitation rate coefficient for the combined levels is the weighted average of the de-excitation rates coefficient for the sublevels
\begin{align}
k_{ul} &= \frac{1}{g_u}\sum_{i,j} g_i \hat{k}_{ij} \\
&= \frac{1}{g_u} \frac{h^2}{2 \pi m_e^2}\left(\frac{m_e}{2 \pi k_B T}\right)^{1/2} \sum_{i,j} \hat{\Upsilon}_{il} \\
&= 8.629\times 10^{-6}\frac{\Upsilon_{ul}}{g_u T^{1/2}} \,\, \rm{cm^3 \, s^{-1}}.
\end{align}
Hence the sum of the collision strengths for sublevel transitions gives the collision strength for the combined level transitions. We fit this combined collision strength with a power-law
\begin{align}
\Upsilon_{ul} = \alpha_0 T^{\beta_0}
\end{align}
which allows us to express the rate coefficients as power-laws, Eq.~\ref{eq:deex}, with the fit coefficients listed in Table~\ref{tab:Hparams}.

\begin{table*}
  \centering
  \caption{Atomic hydrogen sub-level effective collision strengths, $\hat{\Upsilon}_{il}$. Data taken from \cite{anderson_2002} with modifications decribed in the text.}
\begin{tabular}{llllllllll}\cline{1-10}
  \multicolumn{10}{c}{\vspace{-0.3cm}} \\ 
  & & \multicolumn{8}{c}{Energies (eV)} \\ 
i & j & 0.5 & 1.0 & 3.0 & 5.0 & 10.0 & 15.0 & 20.0 & 25.0 \\ \cline{1-10}
  \multicolumn{10}{c}{\vspace{-0.3cm}} \\ 
2 & 1 & $1.43(-01)$ & $1.76(-01)$ & $2.84(-01)$ & $3.83(-01)$ & $6.03(-01)$ & $7.83(-01)$ & $9.37(-01)$ & $1.07(+00)$ \\
3 & 1 & $2.60(-01)$ & $2.96(-01)$ & $3.26(-01)$ & $3.39(-01)$ & $3.73(-01)$ & $4.06(-01)$ & $4.36(-01)$ & $4.61(-01)$ \\
4 & 1 & $2.86(-01)$ & $3.53(-01)$ & $5.69(-01)$ & $7.67(-01)$ & $1.21(+00)$ & $1.57(+00)$ & $1.87(+00)$ & $2.13(+00)$ \\
5 & 1 & $3.73(-02)$ & $4.20(-02)$ & $6.20(-02)$ & $8.10(-02)$ & $1.18(-01)$ & $1.46(-01)$ & $1.69(-01)$ & $1.89(-01)$ \\
5 & 2 & $8.07(-01)$ & $8.80(-01)$ & $1.19(+00)$ & $1.48(+00)$ & $1.97(+00)$ & $2.27(+00)$ & $2.48(+00)$ & $2.63(+00)$ \\
5 & 3 & $8.20(-01)$ & $1.02(+00)$ & $1.76(+00)$ & $2.58(+00)$ & $4.50(+00)$ & $6.13(+00)$ & $7.47(+00)$ & $8.60(+00)$ \\
5 & 4 & $1.61(+00)$ & $1.76(+00)$ & $2.38(+00)$ & $2.96(+00)$ & $3.93(+00)$ & $4.53(+00)$ & $4.96(+00)$ & $5.27(+00)$ \\
6 & 1 & $6.51(-02)$ & $6.96(-02)$ & $7.76(-02)$ & $8.13(-02)$ & $8.70(-02)$ & $9.21(-02)$ & $9.66(-02)$ & $1.01(-01)$ \\
6 & 2 & $6.80(-01)$ & $7.40(-01)$ & $7.60(-01)$ & $7.83(-01)$ & $8.93(-01)$ & $1.01(+00)$ & $1.11(+00)$ & $1.21(+00)$ \\
6 & 3 & $1.38(+00)$ & $1.45(+00)$ & $2.28(+00)$ & $3.09(+00)$ & $4.50(+00)$ & $5.40(+00)$ & $6.03(+00)$ & $6.50(+00)$ \\
6 & 4 & $1.36(+00)$ & $1.48(+00)$ & $1.52(+00)$ & $1.57(+00)$ & $1.79(+00)$ & $2.01(+00)$ & $2.23(+00)$ & $2.41(+00)$ \\
7 & 1 & $2.48(-02)$ & $2.63(-02)$ & $3.13(-02)$ & $3.59(-02)$ & $4.36(-02)$ & $4.80(-02)$ & $5.04(-02)$ & $5.20(-02)$ \\
7 & 2 & $1.76(+00)$ & $2.37(+00)$ & $4.92(+00)$ & $7.44(+00)$ & $1.26(+01)$ & $1.64(+01)$ & $1.95(+01)$ & $2.20(+01)$ \\
7 & 3 & $8.36(-01)$ & $1.23(+00)$ & $2.62(+00)$ & $3.75(+00)$ & $5.68(+00)$ & $6.88(+00)$ & $7.72(+00)$ & $8.32(+00)$ \\
7 & 4 & $3.52(+00)$ & $4.75(+00)$ & $9.84(+00)$ & $1.49(+01)$ & $2.52(+01)$ & $3.28(+01)$ & $3.89(+01)$ & $4.40(+01)$ \\
8 & 1 & $7.47(-02)$ & $8.40(-02)$ & $1.24(-01)$ & $1.62(-01)$ & $2.36(-01)$ & $2.92(-01)$ & $3.38(-01)$ & $3.77(-01)$ \\
8 & 2 & $1.61(+00)$ & $1.76(+00)$ & $2.38(+00)$ & $2.96(+00)$ & $3.93(+00)$ & $4.53(+00)$ & $4.96(+00)$ & $5.27(+00)$ \\
8 & 3 & $1.64(+00)$ & $2.03(+00)$ & $3.52(+00)$ & $5.16(+00)$ & $9.00(+00)$ & $1.23(+01)$ & $1.49(+01)$ & $1.72(+01)$ \\
8 & 4 & $3.23(+00)$ & $3.52(+00)$ & $4.76(+00)$ & $5.91(+00)$ & $7.87(+00)$ & $9.07(+00)$ & $9.91(+00)$ & $1.05(+01)$ \\
9 & 1 & $3.73(-02)$ & $3.95(-02)$ & $4.69(-02)$ & $5.38(-02)$ & $6.54(-02)$ & $7.20(-02)$ & $7.56(-02)$ & $7.80(-02)$ \\
9 & 2 & $2.64(+00)$ & $3.56(+00)$ & $7.38(+00)$ & $1.12(+01)$ & $1.89(+01)$ & $2.46(+01)$ & $2.92(+01)$ & $3.30(+01)$ \\
9 & 3 & $1.25(+00)$ & $1.85(+00)$ & $3.94(+00)$ & $5.63(+00)$ & $8.52(+00)$ & $1.03(+01)$ & $1.16(+01)$ & $1.25(+01)$ \\
9 & 4 & $5.28(+00)$ & $7.12(+00)$ & $1.48(+01)$ & $2.23(+01)$ & $3.78(+01)$ & $4.92(+01)$ & $5.84(+01)$ & $6.60(+01)$ \\ \cline{1-10}
\end{tabular}\label{tab:Hcolstrength}

\emph{Note}. Numbers in parentheses denote powers of ten.
\end{table*}

\section{Two-stream approximation test}\label{app:twostream}
Our ALI transfer code is tested with the two-stream approximation test where the RTE of a homogeneous slab is solved along two rays with $\mu=\pm 1/\sqrt{3}$. In addition, we impose a boundary condition that deep in the slab (when $\tau \to \infty$) the source function equals the Planck function. This setup has analytic solution
\begin{align}
\frac{S}{B} = 1 - \left(1-\epsilon \right)\exp\left( - \tau \sqrt{3\epsilon} \right) \label{eq:twostream-analytic}
\end{align}
where $B$ is the Planck function
\begin{align}
B = \frac{2h\nu^3/c^2}{\exp\left(h\nu / k_B T\right) - 1}
\end{align}
and $\epsilon$ is the non-LTE parameter
\begin{align}
\epsilon = \frac{ C_{ij} } { C_{ij} + A_{ij}/\left(1 - \exp\left(-E_{ij}/k_B T\right) \right)}.
\end{align}
For the homogeneous slab we use the Ly$\alpha$ radiative parameters listed in Table~\ref{tab:Hparams}, and densities and temperatures typical for shock conditions: $T=10^4$~K, $n_e=10^3$~cm$^{-3}$. This gives a non-LTE parameter $\epsilon \sim 10^{-14}$.
  
\begin{figure}
\centering
  \includegraphics[width=\columnwidth]{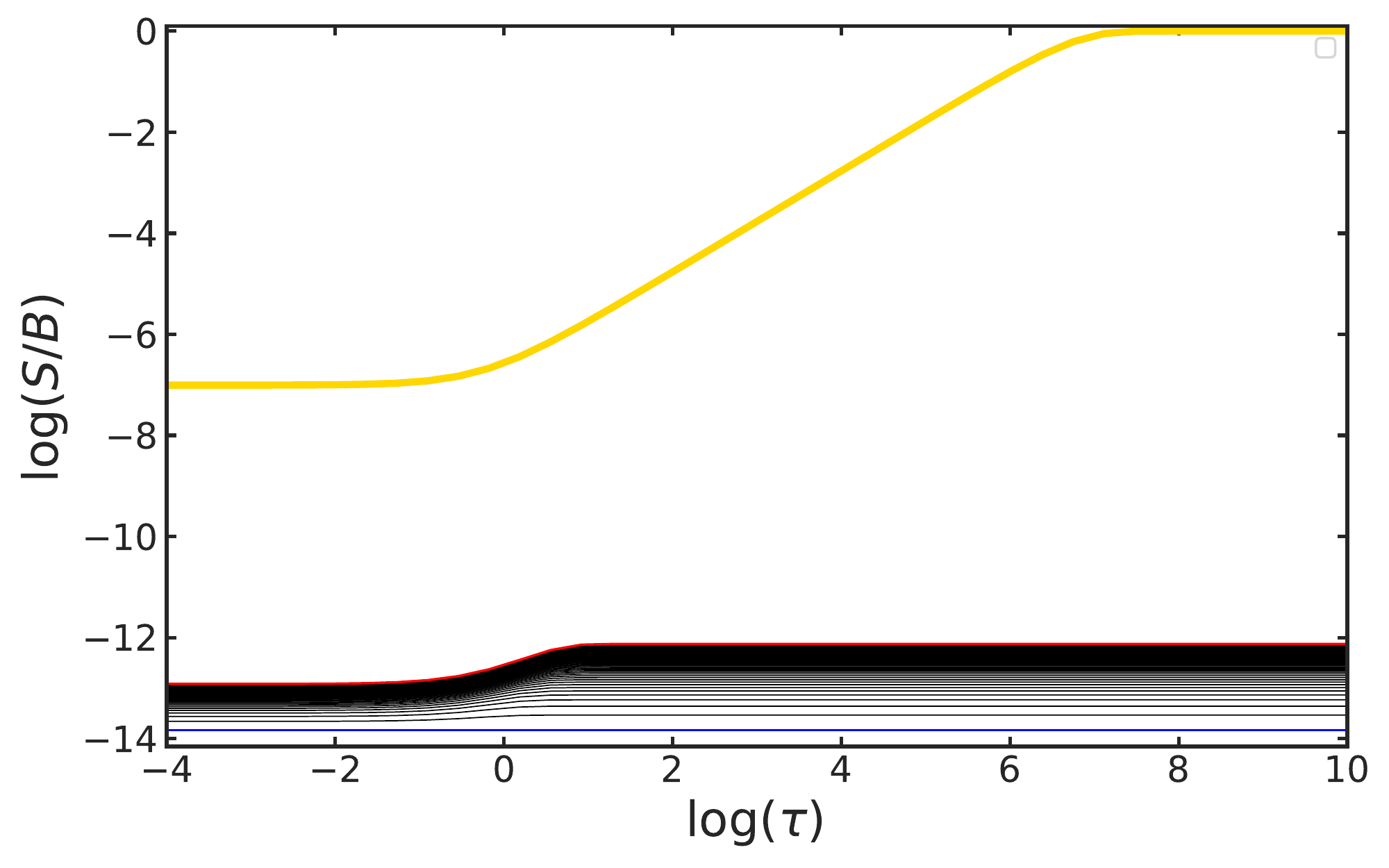}
  \caption{Two-stream approximation test: Lambda Iteration scheme. The yellow line is the analytic solution defined by Eq.~\eqref{eq:twostream-analytic}. The blue line is the initial condition and the red line is the solution after 50 iterations.}
  \label{fig:twostream-noALI}
\end{figure}

\begin{figure}
\centering
  \includegraphics[width=\columnwidth]{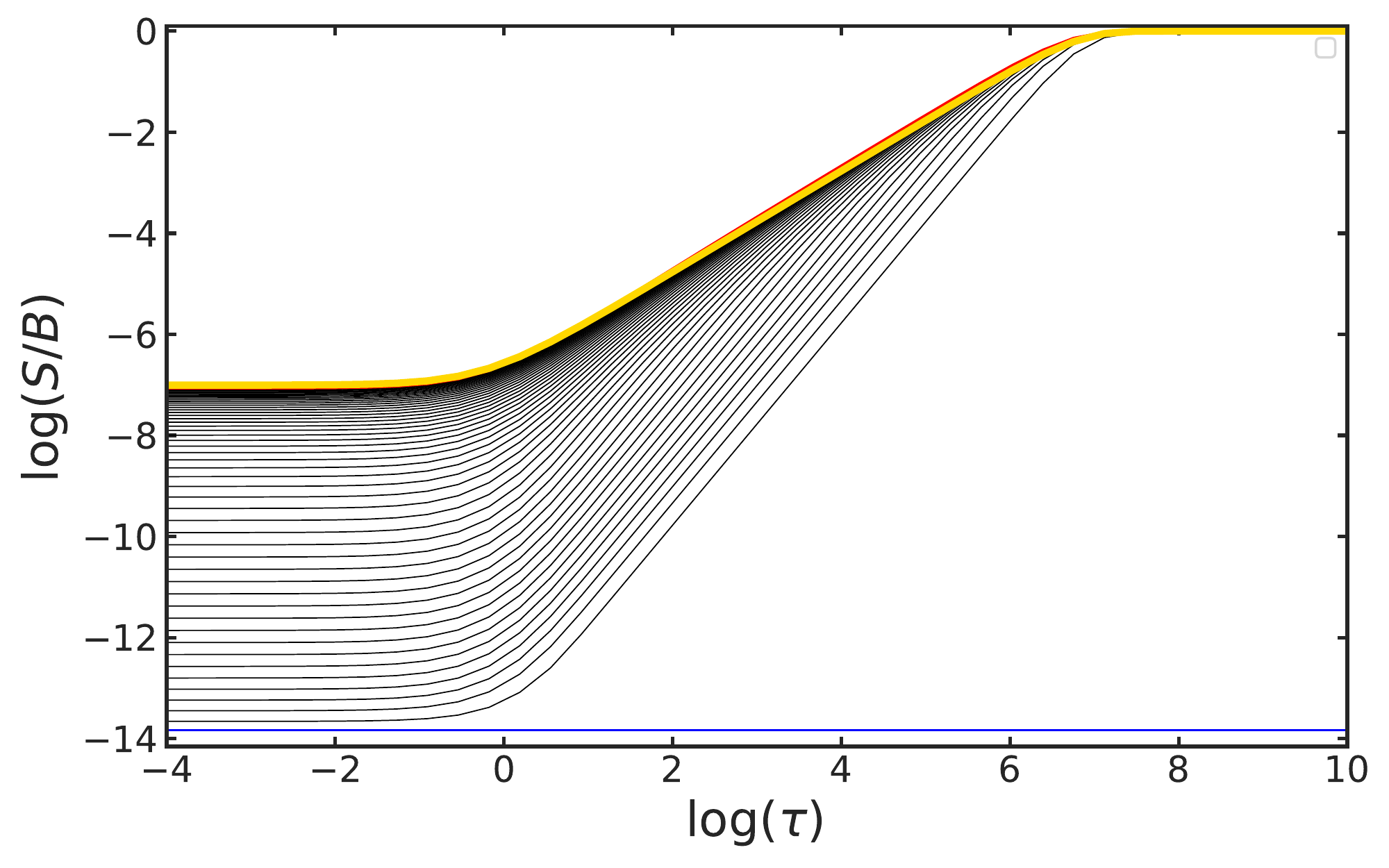}
  \caption{Same as Fig.~\ref{fig:twostream-noALI} but using the Accelerated Lambda Iteration scheme.}
  \label{fig:twostream-ALI}
\end{figure}

\begin{figure}
\centering
  \includegraphics[width=\columnwidth]{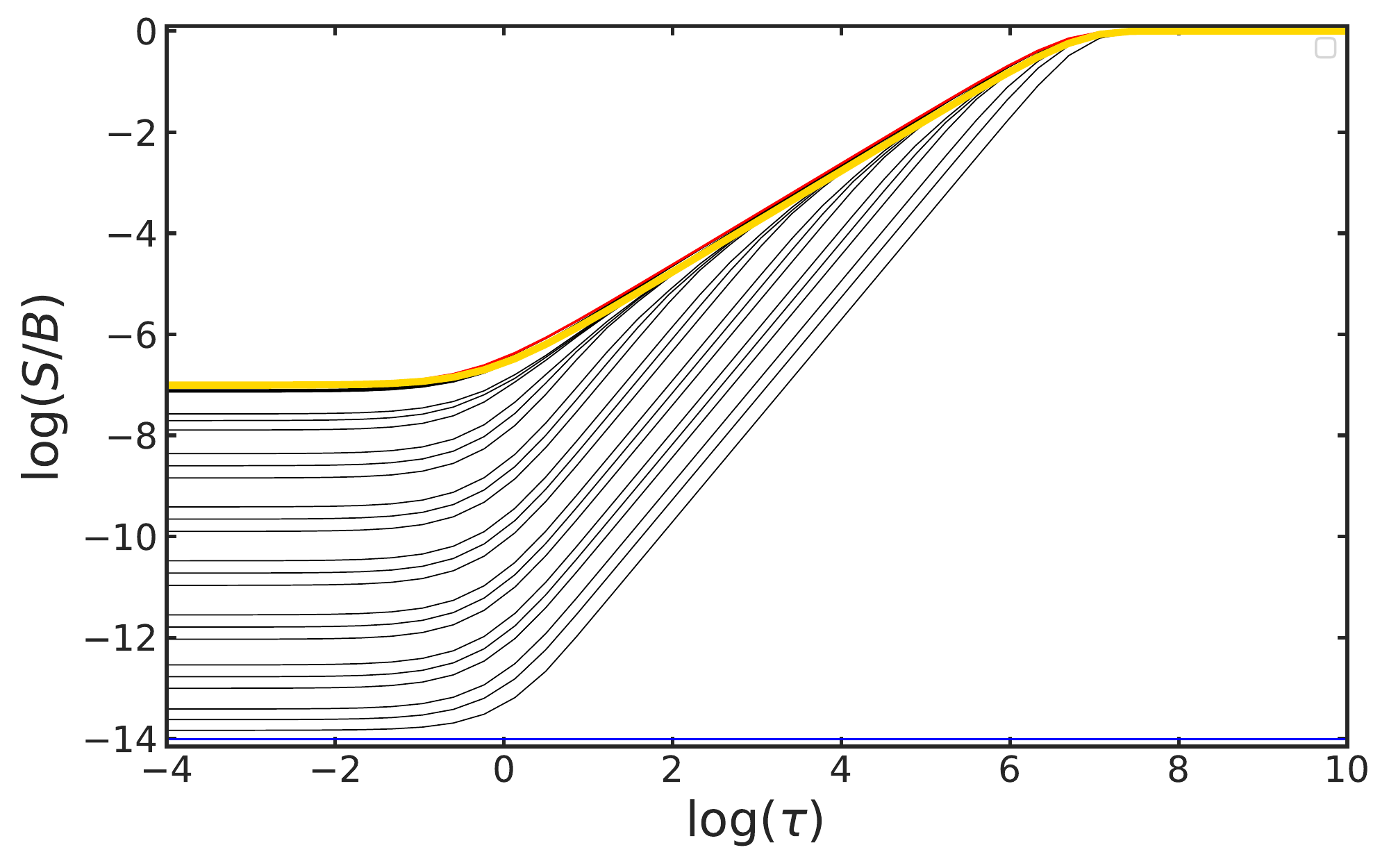}
  \caption{Same as Fig.~\ref{fig:twostream-noALI} but using the Accelerated Lambda Iteration scheme with Ng acceleration. We note that the red line is under the yellow line and cannot be seen.}
  \label{fig:twostream-ALIng}
\end{figure}

The result without using the ALI modification (that is, with $\Lambda^*=0$) is shown in Fig~\ref{fig:twostream-noALI}. After 50 iterations the solution is still more than 6 orders of magnitude below the analytic solution. Using $\Lambda^*$ as defined by Eq.~\eqref{eq:lambdastar} we see in Fig.~\ref{fig:twostream-ALI} that the solution converges onto the analytic solution before 50 iterations. Finally, using the Ng acceleration algorithm we get even faster convergence in Fig.~\ref{fig:twostream-ALIng}.

\section{Three-level hydrogen test}\label{app:threelevel}
  To further test the implementation of our ALI transfer code, we compare
  with the three-level benchmark problem of \cite{1968rla..conf...27A}.
  In this problem the radiative transfer is solved for a plane-parallel
  semi-infinite atmosphere with constant collisional de-excitation rates
  $C_{ij}=10^5$~s$^{-1}$ for all three transitions and temperature $5000$~K.
  The statistical weights are $g_1=2$, $g_2=8$, and $g_3=18$. The transition
  frequencies are $\nu_{21}=2.47\times 10^{15}$~Hz and 
  $\nu_{31}=2.93\times 10^{15}$~Hz. The Einstein coefficients are 
  $A_{21}=4.68\times 10^8$~s$^{-1}$, $A_{31}=5.54\times 10^7$~s$^{-1}$,
  and $A_{32}=4.39\times 10^7$~s$^{-1}$. The line profiles are Gaussian
  (eq.~\ref{eq:lineprofile}), we use an eight-ray Gaussian quadrature
  scheme and the boundary condition
  \begin{align}
  \lim _{\tau \to \infty} S_{ij} = B_{ij}.
  \end{align}
  Finally we initialise the populations to be thermal.
  
\begin{figure}[h]
  \centering
  \includegraphics[width=\columnwidth]{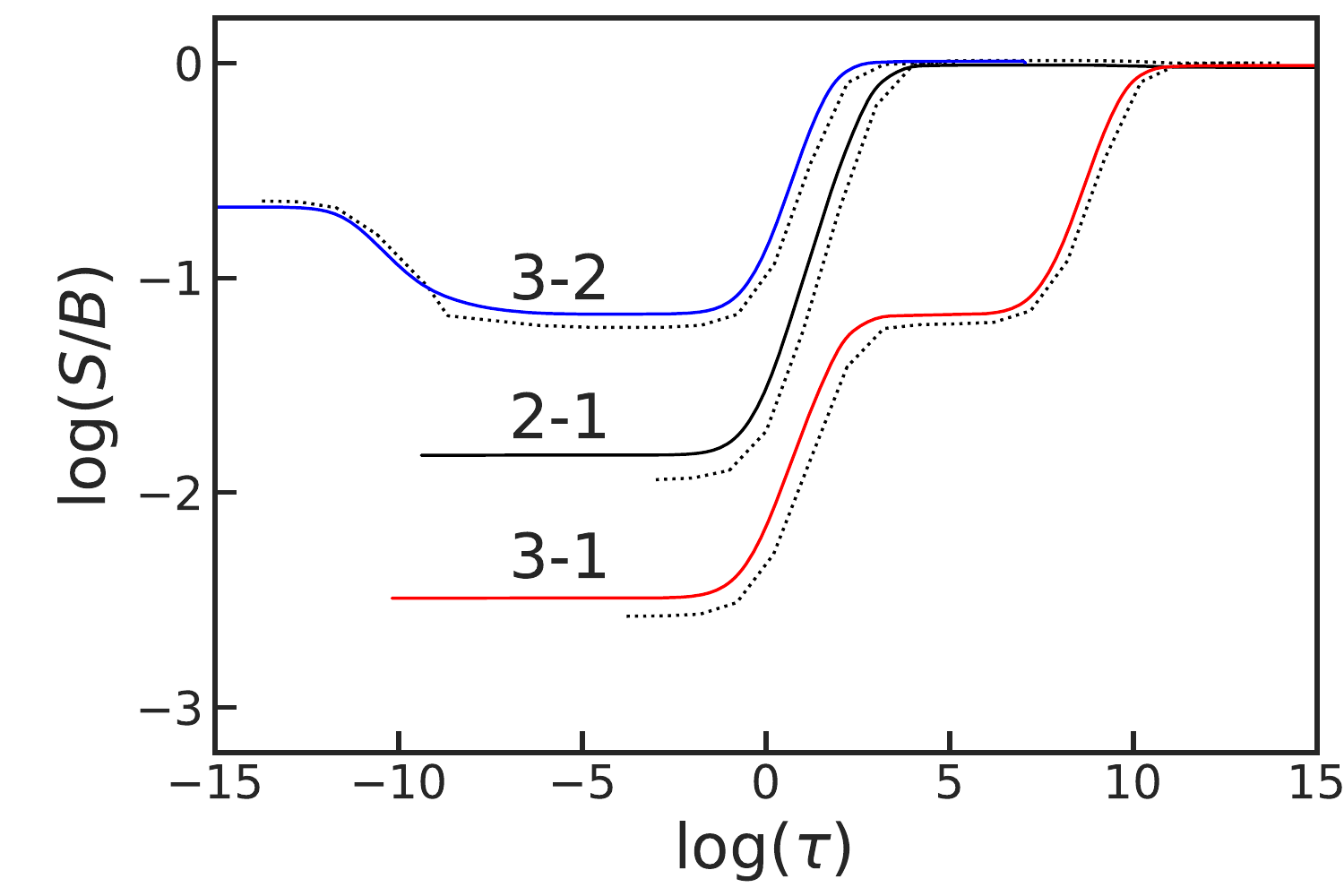}
  \caption{25th iteration of the three-level benchmark problem of \cite{1968rla..conf...27A}. Dotted lines are solutions tabulated in \cite{1968rla..conf...27A}.}
  \label{fig:3lvltest}
\end{figure}

Figure~\ref{fig:3lvltest} shows the ratio of the source to Planck functions for the three transitions as a function of optical depth at the line centers of those transitions after 25 iterations. With $\Lambda^*$ defined by Eq.~\eqref{eq:lambdastar} and implementing Ng acceleration on the source functions the solutions have converged after this many iterations. The solutions show good agreement with the tabulated solutions of \cite{1968rla..conf...27A}.

\section{Tables of results}\label{app:tables}
Here we tabulate results for shock models in the velocity range $V_s=25$--$60$~km/s described in section~\ref{sec:grid}. In table~\ref{tab:intensities} we show the intensities escaping the shock front, which includes material in the shock down to 10~K as discussed in section~\ref{sec:lines}, of hydrogen and atomic fine structure and metastable lines. We also tabulate a subset of H$_2$ rovibrational lines given as output by the code. Additional lines are available upon request. A selection of these lines are shown in figures~\ref{fig:finestructure}--\ref{fig:h2lines}. In table~\ref{tab:columndensities} we give column densities of a variety of species. Some of these column densities are shown in figure~\ref{fig:column_dens1}.

\begin{table*}
  \centering
  \caption{Line and two-photon continuum intensities (${\rm erg/s/cm^2/sr}$) escaping the shock front as a function of shock velocity (km/s) for shocks with preshock density $n_{\rm H}=10^4$~cm$^{-3}$. Emission is integrated only in the postshock region (i.e. without radiative precursor) until the temperature drops to 10~K. The H$_2$ lines are a subset of those output by the code. Additional lines are available upon request.}
\begin{tabular}{l|cccccccc}
 \multicolumn{9}{c}{\vspace{-0.6cm}} \\ 
 & \multicolumn{8}{c}{Velocities (km/s)} \\ 
${\rm Line}$ & 25 & 30 & 35 & 40 & 45 & 50 & 55 & 60 \\ \cline{1-9}
 \multicolumn{9}{c}{\vspace{-0.3cm}} \\ \cline{1-9}
 & \multicolumn{8}{c}{\vspace{-0.3cm}} \\
 & \multicolumn{8}{c}{H lines} \\
${\rm Ly\alpha}\, (1215.7\,\AA)$ & $0$ & $2.96(-04)$ & $3.25(-03)$ & $7.98(-03)$ & $1.41(-02)$ & $2.18(-02)$ & $3.13(-02)$ & $4.36(-02)$ \\
${\rm Ly\beta}\, (1025.7\,\AA)$  & $0$ & $8.14(-05)$ & $1.13(-03)$ & $2.09(-03)$ & $3.11(-03)$ & $4.27(-03)$ & $5.50(-03)$ & $6.98(-03)$ \\ 
${\rm H\alpha}\, (6564.6\,\AA)$  & $0$ & $1.05(-05)$ & $2.37(-04)$ & $5.41(-04)$ & $8.70(-04)$ & $1.27(-03)$ & $1.86(-03)$ & $2.46(-03)$ \\
${\rm 2ph\, (<2400\,\AA)}$      & $0$ & $1.65(-07)$ & $4.03(-05)$ & $2.88(-04)$ & $7.45(-04)$ & $1.42(-03)$ & $2.35(-03)$ & $3.52(-03)$ \\
 \cline{1-9} & \multicolumn{8}{c}{\vspace{-0.3cm}} \\
 & \multicolumn{8}{c}{Atomic fine structure lines} \\
$\rm{C(370.4}\,\mu\rm{m})$ & $2.18(-06)$ & $4.33(-06)$ & $4.70(-06)$ & $5.47(-06)$ & $3.16(-06)$ & $2.32(-06)$ & $1.99(-06)$ & $1.79(-06)$ \\
$\rm{C(609.8}\,\mu\rm{m})$ & $6.45(-07)$ & $7.80(-07)$ & $1.66(-06)$ & $1.86(-06)$ & $7.94(-07)$ & $3.97(-07)$ & $2.86(-07)$ & $2.47(-07)$ \\
$\rm{C}^+(158\,\mu\rm{m})$ & $6.37(-12)$ & $5.23(-09)$ & $3.60(-07)$ & $1.14(-06)$ & $1.97(-06)$ & $3.03(-06)$ & $3.74(-06)$ & $5.33(-06)$ \\
$\rm{O(63.2}\,\mu\rm{m})$ & $2.52(-04)$ & $6.86(-03)$ & $6.41(-03)$ & $6.36(-03)$ & $6.13(-03)$ & $6.13(-03)$ & $6.11(-03)$ & $6.32(-03)$ \\
$\rm{O(145.3}\,\mu\rm{m})$ & $5.54(-06)$ & $1.90(-04)$ & $1.64(-04)$ & $1.59(-04)$ & $1.53(-04)$ & $1.52(-04)$ & $1.51(-04)$ & $1.53(-04)$ \\
$\rm{S(25.2}\,\mu\rm{m})$ & $4.40(-04)$ & $4.91(-03)$ & $4.50(-03)$ & $4.55(-03)$ & $4.52(-03)$ & $4.49(-03)$ & $4.54(-03)$ & $4.64(-03)$ \\
$\rm{Si(68.5}\,\mu\rm{m})$ & $4.50(-08)$ & $1.96(-06)$ & $1.91(-05)$ & $1.14(-05)$ & $1.29(-05)$ & $1.08(-05)$ & $1.01(-05)$ & $6.21(-06)$ \\
$\rm{Si}^+(34.8\,\mu\rm{m})$ & $2.13(-09)$ & $7.77(-07)$ & $3.45(-04)$ & $4.29(-04)$ & $4.30(-04)$ & $4.70(-04)$ & $4.92(-04)$ & $5.60(-04)$ \\
 \cline{1-9} & \multicolumn{8}{c}{\vspace{-0.3cm}} \\
 & \multicolumn{8}{c}{Atomic metastable lines} \\
$\rm{O(6300}\,\AA)$ & $7.40(-08)$ & $2.01(-04)$ & $1.93(-04)$ & $2.99(-04)$ & $4.57(-04)$ & $6.60(-04)$ & $8.92(-04)$ & $1.18(-03)$ \\
$\rm{O(6363}\,\AA)$ & $2.37(-08)$ & $6.43(-05)$ & $6.20(-05)$ & $9.58(-05)$ & $1.46(-04)$ & $2.11(-04)$ & $2.85(-04)$ & $3.77(-04)$ \\
$\rm{S^+(6731}\,\AA)$ & $9.08(-15)$ & $3.03(-12)$ & $3.87(-05)$ & $1.32(-04)$ & $2.19(-04)$ & $2.87(-04)$ & $3.44(-04)$ & $3.88(-04)$ \\
$\rm{S^+(6716}\,\AA)$ & $1.32(-14)$ & $4.29(-12)$ & $3.56(-05)$ & $8.37(-05)$ & $1.22(-04)$ & $1.55(-04)$ & $1.84(-04)$ & $2.11(-04)$ \\
$\rm{C(9850}\,\AA)$ & $1.35(-11)$ & $5.11(-09)$ & $1.94(-04)$ & $4.37(-04)$ & $6.45(-04)$ & $7.70(-04)$ & $7.83(-04)$ & $7.50(-04)$ \\
$\rm{C(9824}\,\AA)$ & $4.56(-12)$ & $1.72(-09)$ & $6.54(-05)$ & $1.47(-04)$ & $2.18(-04)$ & $2.60(-04)$ & $2.64(-04)$ & $2.53(-04)$ \\
$\rm{C^+(2324.7}\,\AA)$ & $1.03(-18)$ & $6.51(-14)$ & $1.09(-07)$ & $2.70(-07)$ & $5.34(-07)$ & $1.40(-06)$ & $5.24(-06)$ & $1.33(-05)$ \\
$\rm{C^+(2323.5}\,\AA)$ & $1.09(-18)$ & $6.94(-14)$ & $1.16(-07)$ & $2.89(-07)$ & $5.71(-07)$ & $1.50(-06)$ & $5.61(-06)$ & $1.42(-05)$ \\
$\rm{C^+(2328.1}\,\AA)$ & $1.21(-18)$ & $7.70(-14)$ & $1.28(-07)$ & $3.20(-07)$ & $6.31(-07)$ & $1.66(-06)$ & $6.20(-06)$ & $1.57(-05)$ \\
$\rm{C^+(2326.9}\,\AA)$ & $3.33(-18)$ & $2.12(-13)$ & $3.55(-07)$ & $8.84(-07)$ & $1.75(-06)$ & $4.59(-06)$ & $1.72(-05)$ & $4.35(-05)$ \\
$\rm{C^+(2325.4}\,\AA)$ & $6.52(-18)$ & $4.20(-13)$ & $7.03(-07)$ & $1.76(-06)$ & $3.47(-06)$ & $9.13(-06)$ & $3.42(-05)$ & $8.65(-05)$ \\
$\rm{N(5200}\,\AA)$ & $1.68(-12)$ & $1.31(-09)$ & $3.24(-05)$ & $3.77(-05)$ & $3.59(-05)$ & $3.16(-05)$ & $2.60(-05)$ & $2.31(-05)$ \\
$\rm{N(5197}\,\AA)$ & $1.12(-12)$ & $8.73(-10)$ & $3.49(-05)$ & $5.19(-05)$ & $5.40(-05)$ & $4.89(-05)$ & $4.00(-05)$ & $3.47(-05)$ \\
$\rm{N^+(6527}\,\AA)$ & $1.68(-22)$ & $2.41(-18)$ & $1.35(-10)$ & $8.45(-10)$ & $4.06(-09)$ & $2.29(-08)$ & $7.83(-08)$ & $1.44(-07)$ \\
$\rm{N^+(6548}\,\AA)$ & $3.16(-19)$ & $4.53(-15)$ & $2.54(-07)$ & $1.59(-06)$ & $7.64(-06)$ & $4.31(-05)$ & $1.47(-04)$ & $2.71(-04)$ \\
$\rm{N^+(6583}\,\AA)$ & $9.31(-19)$ & $1.33(-14)$ & $7.47(-07)$ & $4.68(-06)$ & $2.25(-05)$ & $1.27(-04)$ & $4.34(-04)$ & $7.98(-04)$ \\ 
 \cline{1-9} & \multicolumn{8}{c}{\vspace{-0.3cm}} \\
 & \multicolumn{8}{c}{H$_2$ rovibrational lines} \\
${\rm 0-0 \, S(0) \, 28 \,\mu m}$ & $6.20(-07)$ & $9.72(-07)$ & $1.42(-06)$ & $1.56(-06)$ & $1.65(-06)$ & $1.75(-06)$ & $1.84(-06)$ & $1.96(-06)$ \\
${\rm 0-0 \, S(1) \, 17 \,\mu m}$ & $1.20(-05)$ & $2.05(-05)$ & $3.24(-05)$ & $3.74(-05)$ & $4.03(-05)$ & $4.38(-05)$ & $4.69(-05)$ & $5.16(-05)$ \\
${\rm 0-0 \, S(2) \, 12 \,\mu m}$ & $8.83(-06)$ & $1.37(-05)$ & $2.14(-05)$ & $2.60(-05)$ & $2.81(-05)$ & $3.13(-05)$ & $3.41(-05)$ & $3.93(-05)$ \\
${\rm 0-0 \, S(3) \, 10 \,\mu m}$ & $4.27(-05)$ & $4.49(-05)$ & $6.15(-05)$ & $7.73(-05)$ & $8.33(-05)$ & $9.44(-05)$ & $1.04(-04)$ & $1.27(-04)$ \\
${\rm 0-0 \, S(4) \, 8.0 \,\mu m}$ & $2.77(-05)$ & $2.09(-05)$ & $2.09(-05)$ & $2.39(-05)$ & $2.49(-05)$ & $2.72(-05)$ & $2.92(-05)$ & $3.49(-05)$ \\
${\rm 0-0 \, S(5) \, 6.9 \,\mu m}$ & $1.01(-04)$ & $7.03(-05)$ & $6.71(-05)$ & $7.14(-05)$ & $7.36(-05)$ & $7.72(-05)$ & $8.03(-05)$ & $8.92(-05)$ \\
${\rm 0-0 \, S(6) \, 6.1 \,\mu m}$ & $4.71(-05)$ & $3.29(-05)$ & $3.06(-05)$ & $3.13(-05)$ & $3.24(-05)$ & $3.33(-05)$ & $3.41(-05)$ & $3.53(-05)$ \\
${\rm 0-0 \, S(7) \, 5.5 \,\mu m}$ & $1.39(-04)$ & $9.53(-05)$ & $9.42(-05)$ & $9.70(-05)$ & $1.01(-04)$ & $1.05(-04)$ & $1.08(-04)$ & $1.10(-04)$ \\
${\rm 0-0 \, S(8) \, 5.1 \,\mu m}$ & $5.79(-05)$ & $3.85(-05)$ & $3.67(-05)$ & $3.74(-05)$ & $3.93(-05)$ & $4.07(-05)$ & $4.19(-05)$ & $4.24(-05)$ \\
${\rm 0-0 \, S(9) \, 4.7 \,\mu m}$ & $1.56(-04)$ & $9.99(-05)$ & $1.00(-04)$ & $1.03(-04)$ & $1.09(-04)$ & $1.14(-04)$ & $1.18(-04)$ & $1.20(-04)$ \\
${\rm 1-0 \, S(0) \, 2.22 \,\mu m}$ & $3.42(-05)$ & $1.70(-05)$ & $1.75(-05)$ & $1.86(-05)$ & $1.99(-05)$ & $2.13(-05)$ & $2.24(-05)$ & $2.37(-05)$ \\
${\rm 1-0 \, S(1) \, 2.12 \,\mu m}$ & $1.45(-04)$ & $6.00(-05)$ & $6.24(-05)$ & $6.72(-05)$ & $7.21(-05)$ & $7.73(-05)$ & $8.12(-05)$ & $8.68(-05)$ \\
${\rm 1-0 \, S(2) \, 2.03 \,\mu m}$ & $6.51(-05)$ & $2.77(-05)$ & $2.67(-05)$ & $2.76(-05)$ & $2.91(-05)$ & $3.08(-05)$ & $3.21(-05)$ & $3.38(-05)$ \\
${\rm 1-0 \, S(3) \, 1.96 \,\mu m}$ & $1.82(-04)$ & $7.18(-05)$ & $7.14(-05)$ & $7.54(-05)$ & $8.05(-05)$ & $8.58(-05)$ & $8.98(-05)$ & $9.47(-05)$ \\
${\rm 2-1 \, S(0) \, 2.36 \,\mu m}$ & $8.23(-06)$ & $3.44(-06)$ & $2.88(-06)$ & $2.56(-06)$ & $2.57(-06)$ & $2.68(-06)$ & $2.79(-06)$ & $2.90(-06)$ \\
${\rm 2-1 \, S(1) \, 2.25 \,\mu m}$ & $3.67(-05)$ & $1.25(-05)$ & $1.09(-05)$ & $1.06(-05)$ & $1.10(-05)$ & $1.17(-05)$ & $1.22(-05)$ & $1.29(-05)$ \\
${\rm 2-1 \, S(2) \, 2.15 \,\mu m}$ & $1.94(-05)$ & $8.38(-06)$ & $6.83(-06)$ & $5.73(-06)$ & $5.55(-06)$ & $5.72(-06)$ & $5.91(-06)$ & $6.13(-06)$ \\
${\rm 2-1 \, S(3) \, 2.07 \,\mu m}$ & $5.46(-05)$ & $1.89(-05)$ & $1.58(-05)$ & $1.47(-05)$ & $1.51(-05)$ & $1.59(-05)$ & $1.66(-05)$ & $1.75(-05)$ \\
${\rm 2-0 \, S(0) \, 1.19 \,\mu m}$ & $5.63(-06)$ & $2.35(-06)$ & $1.97(-06)$ & $1.75(-06)$ & $1.76(-06)$ & $1.84(-06)$ & $1.91(-06)$ & $1.98(-06)$ \\
${\rm 2-0 \, S(1) \, 1.16 \,\mu m}$ & $2.71(-05)$ & $9.18(-06)$ & $8.02(-06)$ & $7.78(-06)$ & $8.11(-06)$ & $8.61(-06)$ & $9.00(-06)$ & $9.51(-06)$ \\
${\rm 2-0 \, S(2) \, 1.14 \,\mu m}$ & $1.56(-05)$ & $6.73(-06)$ & $5.49(-06)$ & $4.60(-06)$ & $4.46(-06)$ & $4.59(-06)$ & $4.75(-06)$ & $4.93(-06)$ \\
${\rm 2-0 \, S(3) \, 1.12 \,\mu m}$ & $4.85(-05)$ & $1.68(-05)$ & $1.40(-05)$ & $1.31(-05)$ & $1.34(-05)$ & $1.41(-05)$ & $1.47(-05)$ & $1.55(-05)$ \\
 \cline{1-9}
\end{tabular}\label{tab:intensities}

\emph{Note}. Numbers in parentheses denote powers of ten.
\end{table*}

\begin{table*}
  \centering
  \caption{Column densities (${\rm cm^{-2}}$) of select species warmed above 10~K as a function of shock velocity (km/s) for shocks with preshock density $n_{\rm H}=10^4$~cm$^{-3}$. For each species we show the column densities calculated in the postshock region only (post) and postshock and precursor regions together (post+pre).}
\begin{tabular}{ll|cccccccc}
\multicolumn{9}{c}{\vspace{-0.7cm}} \\
  & & \multicolumn{8}{c}{Velocities (km/s)} \\ 
${\rm Species}$ & Slab & 25 & 30 & 35 & 40 & 45 & 50 & 55 & 60 \\ \cline{1-10}
\multicolumn{9}{c}{\vspace{-0.3cm}} \\ \cline{1-10}
 \cline{1-10} & & \multicolumn{8}{c}{\vspace{-0.3cm}} \\
${\rm H + 2H_2}$ & ${\rm post}$ & $8.1(+20)$ & $1.2(+21)$ & $3.3(+21)$ & $4.2(+21)$ & $4.8(+21)$ & $5.2(+21)$ & $5.6(+21)$ & $6.0(+21)$ \\
 & ${\rm post \, + \, pre}$ & $8.1(+20)$ & $1.2(+21)$ & $9.5(+21)$ & $1.1(+22)$ & $1.2(+22)$ & $1.3(+22)$ & $1.4(+22)$ & $1.5(+22)$ \\
 \cline{1-10} & & \multicolumn{8}{c}{\vspace{-0.3cm}} \\
${\rm H}$ & ${\rm post}$ & $9.8(+19)$ & $3.9(+20)$ & $3.0(+20)$ & $2.8(+20)$ & $2.8(+20)$ & $2.7(+20)$ & $2.7(+20)$ & $2.7(+20)$ \\
 & ${\rm post \, + \, pre}$ & $9.8(+19)$ & $3.9(+20)$ & $3.0(+20)$ & $2.8(+20)$ & $2.8(+20)$ & $2.8(+20)$ & $2.7(+20)$ & $2.8(+20)$ \\
 \cline{1-10} & & \multicolumn{8}{c}{\vspace{-0.3cm}} \\
${\rm H}_2$ & ${\rm post}$ & $3.6(+20)$ & $4.1(+20)$ & $1.5(+21)$ & $2.0(+21)$ & $2.2(+21)$ & $2.4(+21)$ & $2.7(+21)$ & $2.8(+21)$ \\
 & ${\rm post \, + \, pre}$ & $3.6(+20)$ & $4.1(+20)$ & $4.6(+21)$ & $5.4(+21)$ & $6.0(+21)$ & $6.4(+21)$ & $6.8(+21)$ & $7.1(+21)$ \\
 \cline{1-10} & & \multicolumn{8}{c}{\vspace{-0.3cm}} \\
${\rm C}$ & ${\rm post}$ & $8.3(+16)$ & $1.0(+17)$ & $1.9(+17)$ & $1.7(+17)$ & $1.0(+17)$ & $5.3(+16)$ & $3.6(+16)$ & $2.8(+16)$ \\
 & ${\rm post \, + \, pre}$ & $8.3(+16)$ & $1.0(+17)$ & $2.0(+17)$ & $1.9(+17)$ & $1.2(+17)$ & $6.9(+16)$ & $5.1(+16)$ & $4.3(+16)$ \\
 \cline{1-10} & & \multicolumn{8}{c}{\vspace{-0.3cm}} \\
${\rm C}^+$ & ${\rm post}$ & $4.2(+09)$ & $3.5(+12)$ & $1.8(+14)$ & $7.1(+14)$ & $1.3(+15)$ & $2.0(+15)$ & $2.7(+15)$ & $3.5(+15)$ \\
 & ${\rm post \, + \, pre}$ & $4.2(+09)$ & $3.5(+12)$ & $1.2(+15)$ & $5.6(+15)$ & $8.7(+15)$ & $1.1(+16)$ & $1.3(+16)$ & $1.5(+16)$ \\
 \cline{1-10} & & \multicolumn{8}{c}{\vspace{-0.3cm}} \\
${\rm O}$ & ${\rm post}$ & $2.5(+16)$ & $2.5(+17)$ & $7.3(+17)$ & $8.6(+17)$ & $8.9(+17)$ & $9.0(+17)$ & $9.6(+17)$ & $1.0(+18)$ \\
 & ${\rm post \, + \, pre}$ & $2.5(+16)$ & $2.5(+17)$ & $1.7(+18)$ & $2.0(+18)$ & $2.1(+18)$ & $2.2(+18)$ & $2.3(+18)$ & $2.4(+18)$ \\
 \cline{1-10} & & \multicolumn{8}{c}{\vspace{-0.3cm}} \\
${\rm S}$ & ${\rm post}$ & $1.5(+16)$ & $2.3(+16)$ & $6.2(+16)$ & $7.8(+16)$ & $8.8(+16)$ & $9.5(+16)$ & $1.0(+17)$ & $1.1(+17)$ \\
 & ${\rm post \, + \, pre}$ & $1.5(+16)$ & $2.3(+16)$ & $1.6(+17)$ & $1.7(+17)$ & $1.8(+17)$ & $1.9(+17)$ & $2.0(+17)$ & $2.0(+17)$ \\
 \cline{1-10} & & \multicolumn{8}{c}{\vspace{-0.3cm}} \\
${\rm S}^+$ & ${\rm post}$ & $6.6(+10)$ & $1.3(+12)$ & $1.2(+14)$ & $5.0(+14)$ & $9.4(+14)$ & $1.4(+15)$ & $1.8(+15)$ & $2.3(+15)$ \\
 & ${\rm post \, + \, pre}$ & $6.6(+10)$ & $1.3(+12)$ & $1.2(+16)$ & $3.4(+16)$ & $4.6(+16)$ & $5.5(+16)$ & $6.2(+16)$ & $6.8(+16)$ \\
 \cline{1-10} & & \multicolumn{8}{c}{\vspace{-0.3cm}} \\
${\rm Si}$ & ${\rm post}$ & $1.9(+13)$ & $8.8(+13)$ & $8.5(+15)$ & $1.0(+16)$ & $1.1(+16)$ & $1.2(+16)$ & $1.3(+16)$ & $1.3(+16)$ \\
 & ${\rm post \, + \, pre}$ & $1.9(+13)$ & $8.8(+13)$ & $1.2(+16)$ & $1.3(+16)$ & $1.4(+16)$ & $1.5(+16)$ & $1.5(+16)$ & $1.6(+16)$ \\
 \cline{1-10} & & \multicolumn{8}{c}{\vspace{-0.3cm}} \\
${\rm Si}^+$ & ${\rm post}$ & $4.7(+09)$ & $2.4(+12)$ & $1.5(+15)$ & $2.3(+15)$ & $3.1(+15)$ & $3.7(+15)$ & $4.2(+15)$ & $4.7(+15)$ \\
 & ${\rm post \, + \, pre}$ & $4.7(+09)$ & $2.4(+12)$ & $1.6(+16)$ & $1.9(+16)$ & $2.2(+16)$ & $2.4(+16)$ & $2.6(+16)$ & $2.7(+16)$ \\
 \cline{1-10} & & \multicolumn{8}{c}{\vspace{-0.3cm}} \\
${\rm SiO}$ & ${\rm post}$ & $2.4(+15)$ & $3.5(+15)$ & $2.2(+13)$ & $1.3(+13)$ & $9.6(+12)$ & $8.1(+12)$ & $7.5(+12)$ & $7.0(+12)$ \\
 & ${\rm post \, + \, pre}$ & $2.4(+15)$ & $3.5(+15)$ & $7.3(+14)$ & $6.8(+14)$ & $6.7(+14)$ & $6.7(+14)$ & $6.7(+14)$ & $6.9(+14)$ \\
 \cline{1-10} & & \multicolumn{8}{c}{\vspace{-0.3cm}} \\
${\rm SH}$ & ${\rm post}$ & $1.6(+12)$ & $6.6(+11)$ & $3.2(+11)$ & $1.8(+11)$ & $1.6(+11)$ & $1.6(+11)$ & $1.5(+11)$ & $1.5(+11)$ \\
 & ${\rm post \, + \, pre}$ & $1.6(+12)$ & $6.6(+11)$ & $8.4(+12)$ & $8.6(+12)$ & $9.0(+12)$ & $9.4(+12)$ & $9.8(+12)$ & $1.0(+13)$ \\
 \cline{1-10} & & \multicolumn{8}{c}{\vspace{-0.3cm}} \\
${\rm SH}^+$ & ${\rm post}$ & $8.2(+08)$ & $8.1(+08)$ & $1.1(+11)$ & $4.5(+10)$ & $3.4(+10)$ & $4.0(+10)$ & $4.8(+10)$ & $5.6(+10)$ \\
 & ${\rm post \, + \, pre}$ & $8.2(+08)$ & $8.1(+08)$ & $3.7(+11)$ & $2.7(+11)$ & $2.6(+11)$ & $2.6(+11)$ & $2.7(+11)$ & $2.8(+11)$ \\
 \cline{1-10} & & \multicolumn{8}{c}{\vspace{-0.3cm}} \\
${\rm CO}$ & ${\rm post}$ & $2.0(+16)$ & $4.7(+16)$ & $2.7(+17)$ & $4.1(+17)$ & $5.5(+17)$ & $6.6(+17)$ & $7.3(+17)$ & $7.9(+17)$ \\
 & ${\rm post \, + \, pre}$ & $2.0(+16)$ & $4.7(+16)$ & $1.1(+18)$ & $1.3(+18)$ & $1.6(+18)$ & $1.7(+18)$ & $1.8(+18)$ & $1.9(+18)$ \\
 \cline{1-10} & & \multicolumn{8}{c}{\vspace{-0.3cm}} \\
${\rm CH}$ & ${\rm post}$ & $1.6(+13)$ & $2.7(+12)$ & $2.4(+12)$ & $1.2(+13)$ & $2.4(+13)$ & $2.7(+13)$ & $2.3(+13)$ & $1.8(+13)$ \\
 & ${\rm post \, + \, pre}$ & $1.6(+13)$ & $2.7(+12)$ & $7.1(+12)$ & $1.7(+13)$ & $3.0(+13)$ & $3.3(+13)$ & $2.9(+13)$ & $2.4(+13)$ \\
 \cline{1-10} & & \multicolumn{8}{c}{\vspace{-0.3cm}} \\
${\rm CH}^+$ & ${\rm post}$ & $2.5(+09)$ & $8.6(+08)$ & $1.4(+11)$ & $3.3(+11)$ & $7.4(+11)$ & $1.4(+12)$ & $2.1(+12)$ & $2.9(+12)$ \\
 & ${\rm post \, + \, pre}$ & $2.5(+09)$ & $8.6(+08)$ & $1.4(+11)$ & $3.3(+11)$ & $7.4(+11)$ & $1.4(+12)$ & $2.1(+12)$ & $3.0(+12)$ \\
 \cline{1-10} & & \multicolumn{8}{c}{\vspace{-0.3cm}} \\
${\rm O}_2$ & ${\rm post}$ & $6.2(+12)$ & $1.3(+14)$ & $1.4(+14)$ & $1.0(+14)$ & $8.0(+13)$ & $6.6(+13)$ & $5.7(+13)$ & $5.0(+13)$ \\
 & ${\rm post \, + \, pre}$ & $6.2(+12)$ & $1.3(+14)$ & $1.7(+16)$ & $1.6(+16)$ & $1.6(+16)$ & $1.7(+16)$ & $1.7(+16)$ & $1.7(+16)$ \\
 \cline{1-10} & & \multicolumn{8}{c}{\vspace{-0.3cm}} \\
${\rm OH}$ & ${\rm post}$ & $1.4(+14)$ & $1.6(+14)$ & $2.5(+14)$ & $2.9(+14)$ & $3.1(+14)$ & $3.4(+14)$ & $3.6(+14)$ & $3.7(+14)$ \\
 & ${\rm post \, + \, pre}$ & $1.4(+14)$ & $1.6(+14)$ & $1.1(+15)$ & $1.2(+15)$ & $1.2(+15)$ & $1.3(+15)$ & $1.4(+15)$ & $1.5(+15)$ \\
 \cline{1-10} & & \multicolumn{8}{c}{\vspace{-0.3cm}} \\
${\rm OH}^+$ & ${\rm post}$ & $1.2(+07)$ & $7.7(+06)$ & $4.6(+10)$ & $8.1(+10)$ & $7.7(+10)$ & $5.8(+10)$ & $3.8(+10)$ & $2.4(+10)$ \\
 & ${\rm post \, + \, pre}$ & $1.2(+07)$ & $7.7(+06)$ & $4.9(+10)$ & $8.3(+10)$ & $7.9(+10)$ & $6.1(+10)$ & $4.1(+10)$ & $2.7(+10)$ \\
 \cline{1-10} & & \multicolumn{8}{c}{\vspace{-0.3cm}} \\
${\rm H_2O}$ & ${\rm post}$ & $2.0(+17)$ & $6.7(+16)$ & $2.7(+14)$ & $1.9(+14)$ & $1.5(+14)$ & $1.4(+14)$ & $1.3(+14)$ & $1.2(+14)$ \\
 & ${\rm post \, + \, pre}$ & $2.0(+17)$ & $6.7(+16)$ & $1.5(+15)$ & $1.4(+15)$ & $1.4(+15)$ & $1.4(+15)$ & $1.4(+15)$ & $1.5(+15)$ \\
 \cline{1-10} & & \multicolumn{8}{c}{\vspace{-0.3cm}} \\
${\rm H_2S}$ & ${\rm post}$ & $4.0(+11)$ & $2.6(+11)$ & $7.4(+10)$ & $4.7(+09)$ & $2.3(+09)$ & $2.4(+09)$ & $2.7(+09)$ & $3.1(+09)$ \\
 & ${\rm post \, + \, pre}$ & $4.0(+11)$ & $2.6(+11)$ & $5.5(+12)$ & $5.2(+12)$ & $5.3(+12)$ & $5.3(+12)$ & $5.5(+12)$ & $5.7(+12)$ \\
 \cline{1-10} & & \multicolumn{8}{c}{\vspace{-0.3cm}} \\
${\rm SO}$ & ${\rm post}$ & $1.8(+12)$ & $6.1(+12)$ & $5.1(+12)$ & $3.7(+12)$ & $2.9(+12)$ & $2.4(+12)$ & $2.1(+12)$ & $1.9(+12)$ \\
 & ${\rm post \, + \, pre}$ & $1.8(+12)$ & $6.1(+12)$ & $5.0(+14)$ & $4.8(+14)$ & $4.8(+14)$ & $4.9(+14)$ & $5.0(+14)$ & $5.2(+14)$ \\
 \cline{1-10} & & \multicolumn{8}{c}{\vspace{-0.3cm}} \\
${\rm SO}_2$ & ${\rm post}$ & $2.2(+13)$ & $5.3(+13)$ & $9.0(+10)$ & $3.6(+10)$ & $1.8(+10)$ & $1.1(+10)$ & $8.5(+09)$ & $6.2(+09)$ \\
 & ${\rm post \, + \, pre}$ & $2.2(+13)$ & $5.3(+13)$ & $5.7(+14)$ & $5.5(+14)$ & $5.5(+14)$ & $5.6(+14)$ & $5.8(+14)$ & $6.0(+14)$ \\
 \cline{1-10} & & \multicolumn{8}{c}{\vspace{-0.3cm}} \\
${\rm CS}$ & ${\rm post}$ & $4.7(+13)$ & $2.5(+13)$ & $4.5(+12)$ & $3.0(+12)$ & $2.1(+12)$ & $1.5(+12)$ & $1.2(+12)$ & $9.0(+11)$ \\
 & ${\rm post \, + \, pre}$ & $4.7(+13)$ & $2.5(+13)$ & $2.1(+14)$ & $2.1(+14)$ & $2.1(+14)$ & $2.1(+14)$ & $2.1(+14)$ & $2.2(+14)$ \\
 \cline{1-10} & & \multicolumn{8}{c}{\vspace{-0.3cm}} \\
${\rm CN}$ & ${\rm post}$ & $2.3(+14)$ & $2.6(+13)$ & $1.0(+14)$ & $9.7(+13)$ & $8.6(+13)$ & $6.9(+13)$ & $5.4(+13)$ & $4.1(+13)$ \\
 & ${\rm post \, + \, pre}$ & $2.3(+14)$ & $2.6(+13)$ & $6.8(+15)$ & $6.7(+15)$ & $6.9(+15)$ & $7.2(+15)$ & $7.3(+15)$ & $7.4(+15)$ \\
 \cline{1-10} & & \multicolumn{8}{c}{\vspace{-0.3cm}} \\
${\rm HCN}$ & ${\rm post}$ & $8.4(+15)$ & $1.6(+16)$ & $4.3(+14)$ & $2.3(+14)$ & $1.4(+14)$ & $9.7(+13)$ & $7.4(+13)$ & $5.5(+13)$ \\
 & ${\rm post \, + \, pre}$ & $8.4(+15)$ & $1.6(+16)$ & $6.5(+14)$ & $4.5(+14)$ & $3.6(+14)$ & $3.2(+14)$ & $3.1(+14)$ & $2.9(+14)$ \\
 \cline{1-10} & & \multicolumn{8}{c}{\vspace{-0.3cm}} \\
${\rm HCO}^+$ & ${\rm post}$ & $2.9(+09)$ & $6.9(+09)$ & $8.8(+10)$ & $3.3(+11)$ & $5.1(+11)$ & $6.1(+11)$ & $6.4(+11)$ & $6.2(+11)$ \\
 & ${\rm post \, + \, pre}$ & $2.9(+09)$ & $6.9(+09)$ & $7.7(+11)$ & $9.5(+11)$ & $1.1(+12)$ & $1.2(+12)$ & $1.3(+12)$ & $1.3(+12)$ \\
 \cline{1-10} & & \multicolumn{8}{c}{\vspace{-0.3cm}} \\
${\rm NH}_3$ & ${\rm post}$ & $2.2(+12)$ & $2.0(+12)$ & $4.9(+11)$ & $6.4(+10)$ & $2.5(+10)$ & $1.8(+10)$ & $1.5(+10)$ & $1.3(+10)$ \\
 & ${\rm post \, + \, pre}$ & $2.2(+12)$ & $2.0(+12)$ & $8.0(+11)$ & $3.6(+11)$ & $3.2(+11)$ & $3.2(+11)$ & $3.3(+11)$ & $3.3(+11)$ \\
 \cline{1-10}
\end{tabular}\label{tab:columndensities}

\emph{Note}. Numbers in parentheses denote powers of ten.
\end{table*}

\section{Comparison with \cite{neufeld_fast_1989}}\label{app:neufeld_comparison}
In Table~\ref{tab:neufeld} we give the column densities of several species obtained with the Paris-Durham shock code for 60~km/s shocks at three densities, $n_{\rm{H}}=10^4$, $10^5$, and $10^6$~cm$^{-3}$ in order to compare with the shock models of \cite{neufeld_fast_1989}. Our column densities are in rough agreement. The differences are due to H$_2$ abundance, where \citeauthor{neufeld_fast_1989} find that cooling has reduced the temperature below 200~K before H$_2$ can reform to more than a few percent. In our shock the gas is fully molecular in the reformation plateau above 200~K. This means the largest differences are in species influenced by the presence of H$_2$. However, after three decades of development the two works are different. There are different computational methods, expanded chemical reaction networks and the inclusion of the magnetic field in our models. Hence it is striking how similar some of the column densities are with only a few species---for instance HCN, SO$_2$, and SiO---varying by $\sim$3 orders of magnitude or more in some models.

\begin{table*}
\centering
\caption{Column densities (cm$^{-2}$) of warm gas, $T > 200$~K, predicted with the Paris-Durham shock code for shocks with velocity $V_s=60$~km/s and ambient densities $n_{\rm H}=10^4$, $10^5$, and $10^6$~cm$^{-3}$ (L20), and comparison with the column densities predicted by \cite{neufeld_fast_1989} (ND89). Total column densities for ND89 are estimated from their figures~14 and 15.}
\begin{tabular}{l|cc|cc|cc}
Species & \multicolumn{2}{c|}{$n_{\rm{H}}=10^4$~cm$^{-3}$} & \multicolumn{2}{c|}{$n_{\rm{H}}=10^5$~cm$^{-3}$} & \multicolumn{2}{c}{$n_{\rm{H}}=10^6$~cm$^{-3}$} \\
 & L20 & ND89 & L20 & ND89 & L20 & ND89 \\ \cline{1-7}
\multicolumn{7}{c}{\vspace{-0.3cm}} \\ 
${\rm H + 2H_2}$ & $4.2(+20)$ & $1.0(+20)$ & $9.2(+20)$ & -- & $1.5(+21)$ & $8.0(+21)$ \\
\cline{1-7} \multicolumn{7}{c}{\vspace{-0.3cm}} \\
O & $9.8(+16)$ & $6.0(+16)$ & $2.0(+17)$ & $1.0(+18)$ & $3.1(+17)$ & $2.5(+18)$ \\
C & $2.7(+16)$ & $1.9(+16)$ & $5.1(+16)$ & $1.3(+17)$ & $6.4(+16)$ & $1.4(+17)$ \\
C$^+$ & $3.5(+15)$ & $3.9(+15)$ & $4.8(+15)$ & $5.3(+15)$ & $1.1(+16)$ & $1.4(+16)$ \\
Si$^+$ & $1.1(+15)$ & $1.3(+15)$ & $1.9(+15)$ & $1.2(+16)$ & $3.3(+15)$ & $5.4(+15)$ \\
S & $6.0(+15)$ & $7.6(+14)$ & $1.5(+16)$ & $2.8(+16)$ & $2.3(+16)$ & $9.5(+16)$ \\
S$^+$ & $1.7(+15)$ & $7.6(+14)$ & $2.3(+15)$ & $7.9(+15)$ & $5.0(+15)$ & $3.6(+15)$ \\ \\
H$_2$ & $7.7(+19)$ & $1.5(+18)$ & $2.0(+20)$ & $6.8(+20)$ & $4.1(+20)$ & $2.5(+21)$ \\
OH & $3.3(+14)$ & $3.4(+14)$ & $1.4(+15)$ & $5.7(+15)$ & $2.3(+15)$ & $2.7(+16)$ \\
H$_2$O & $1.5(+14)$ & $5.8(+12)$ & $2.3(+15)$ & $1.9(+15)$ & $2.5(+15)$ & $6.1(+16)$ \\
CO & $2.8(+16)$ & $2.3(+15)$ & $6.9(+16)$ & $4.6(+17)$ & $1.3(+17)$ & $1.5(+18)$ \\ \\
SiO & $4.4(+12)$ & $3.7(+12)$ & $7.3(+13)$ & $1.8(+16)$ & $6.6(+13)$ & $7.8(+16)$ \\
SO & $1.5(+12)$ & $5.7(+10)$ & $2.8(+13)$ & $2.3(+13)$ & $2.5(+13)$ & $8.6(+14)$ \\
NO & $2.7(+13)$ & $6.9(+11)$ & $3.9(+14)$ & $2.3(+14)$ & $4.5(+14)$ & $3.1(+15)$ \\
CN & $1.2(+13)$ & $2.5(+12)$ & $1.5(+14)$ & $1.9(+14)$ & $8.5(+13)$ & $8.2(+14)$ \\
HCN & $2.1(+13)$ & $5.9(+10)$ & $1.0(+15)$ & $5.0(+14)$ & $3.7(+14)$ & $8.2(+15)$ \\
N$_2$ & $7.9(+12)$ & $2.3(+12)$ & $3.9(+14)$ & $7.6(+15)$ & $2.4(+14)$ & $1.5(+17)$ \\ \\
CO$_2$ & $7.5(+10)$ & $4.1(+09)$ & $1.5(+12)$ & $2.9(+13)$ & $1.4(+12)$ & $1.2(+15)$ \\
O$_2$ & $4.9(+13)$ & $1.4(+12)$ & $5.5(+14)$ & $2.9(+14)$ & $5.0(+14)$ & $7.6(+15)$ \\
SO$_2$ & $4.5(+09)$ & $4.4(+06)$ & $6.2(+11)$ & $9.8(+09)$ & $3.2(+11)$ & $4.6(+12)$ \\ \\
CH & $1.1(+13)$ & $6.3(+11)$ & $5.1(+11)$ & $8.8(+11)$ & $6.1(+11)$ & $1.3(+12)$ \\
CH$_2$ & $1.1(+11)$ & $2.1(+09)$ & $4.7(+09)$ & $8.9(+11)$ & $1.1(+10)$ & $2.2(+11)$ \\
C$_2$ & $9.6(+09)$ & $8.2(+09)$ & $5.1(+09)$ & $6.6(+12)$ & $1.6(+10)$ & $1.0(+13)$ \\
C$_2$H & $1.0(+10)$ & $1.4(+09)$ & $4.0(+09)$ & $4.5(+10)$ & $1.0(+10)$ & $3.8(+10)$ \\
C$_2$H$_2$ & $9.3(+09)$ & $1.6(+07)$ & $2.2(+09)$ & $1.0(+10)$ & $2.6(+09)$ & $1.6(+10)$ \\
C$_3$ & $1.2(+07)$ & $1.9(+05)$ & $5.5(+06)$ & $1.6(+08)$ & $1.7(+07)$ & $5.2(+08)$ \\
C$_3$H & $3.5(+07)$ & $9.3(+04)$ & $1.0(+07)$ & $8.5(+07)$ & $2.6(+07)$ & $3.0(+08)$ \\
C$_3$H$_2$ & $2.9(+06)$ & $2.9(+02)$ & $2.2(+06)$ & $9.8(+05)$ & $3.2(+06)$ & $3.8(+06)$ \\ \\
SH & $1.2(+11)$ & $1.0(+11)$ & $4.9(+11)$ & $1.7(+11)$ & $4.9(+11)$ & $3.3(+11)$ \\
H$_2$S & $1.3(+07)$ & $1.3(+07)$ & $6.1(+07)$ & $2.2(+07)$ & $1.6(+08)$ & $3.2(+07)$ \\
H$_2$CO & $6.8(+10)$ & $6.7(+03)$ & $1.4(+11)$ & $9.8(+06)$ & $1.4(+11)$ & $9.6(+06)$ \\ \\
H$_2^+$ & $2.6(+13)$ & $4.9(+12)$ & $2.6(+13)$ & $5.7(+12)$ & $2.6(+13)$ & $6.5(+12)$ \\
H$_3^+$ & $4.4(+12)$ & $8.5(+11)$ & $4.4(+12)$ & $2.7(+11)$ & $4.4(+12)$ & $9.7(+10)$ \\
OH$^+$ & $2.7(+10)$ & $1.9(+13)$ & $2.3(+10)$ & $1.8(+13)$ & $2.4(+10)$ & $1.5(+13)$ \\
CH$^+$ & $3.3(+12)$ & $1.2(+11)$ & $1.3(+12)$ & $2.9(+11)$ & $4.3(+12)$ & $3.3(+11)$ \\
HCO$^+$ & $7.5(+11)$ & $6.7(+10)$ & $3.5(+11)$ & $6.9(+12)$ & $5.4(+11)$ & $1.1(+13)$ \\
SO$^+$ & $5.2(+11)$ & $3.8(+10)$ & $1.6(+12)$ & $5.6(+12)$ & $2.9(+12)$ & $1.8(+13)$ \\
\cline{1-7}
\end{tabular}\label{tab:neufeld}

\emph{Note}. Numbers in parentheses denote powers of ten.
\end{table*}

\section{Energetics decomposition}\label{app:rosace}
\begin{figure*}
  \centering
  \includegraphics[width=\textwidth]{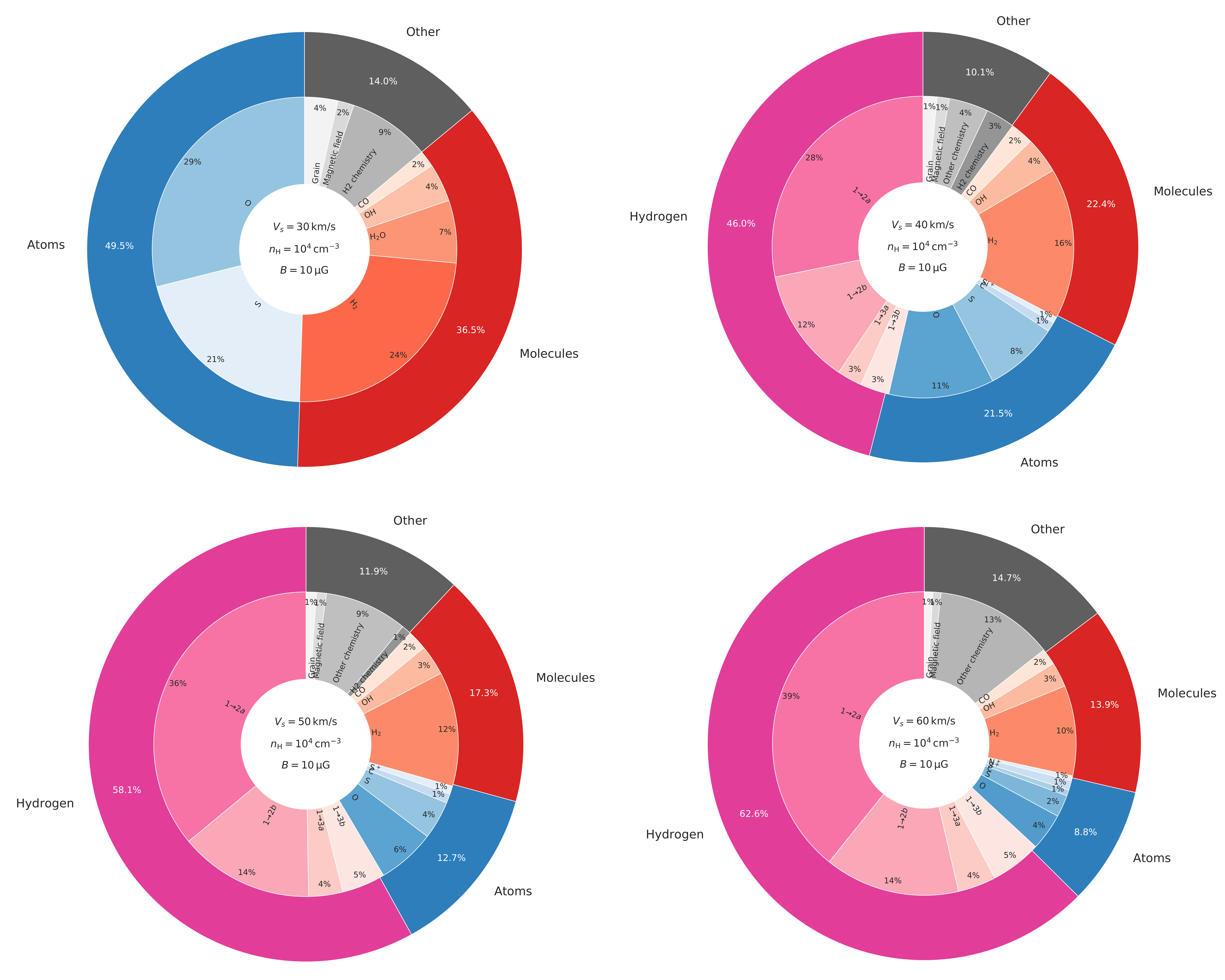}
  \caption{Pathways of energy reprocessing for shocks with velocities 30, 40, 50, and 60~km/s and preshock density $n_{\rm H}=10^4$~cm$^{-3}$. This shows the energy lost due to excitation of atomic H, other atoms, molecules, and other processes as a percentage of the total energy flux. H$_2$ chemistry involves cooling due to collisional dissociation and heating due to reformation. Other chemistry is mostly cooling due to collisional ionisation.}
  \label{fig:rosace_range}
\end{figure*}

In Fig.~\ref{fig:rosace_range} we show the energy reprocessing pathways for shocks with velocities $V_s=30$, 40, 50, and 60~km/s propagating into a medium with density $n_{\rm H}=10^4$~cm$^{-3}$. These figures clearly emphasise that the cooling due to excitation of atomic H becomes suddenly dominant between velocities $V_s=30$ and 40~km/s. They also show that non-H atomic and molecular cooling remain roughly equally important over the whole velocity range.

\end{document}